\documentclass[%
 reprint,
 amsmath,amssymb,
 aps,
]{revtex4-2}

\usepackage{graphicx}
\usepackage{dcolumn}
\usepackage{bm}
\usepackage{xfrac}
\usepackage{upgreek}
\counterwithout{figure}{section}
\usepackage[hidelinks]{hyperref}

\usepackage{mathtools}
\mathtoolsset{centercolon}  

\usepackage{xcolor}

\renewcommand{\phi}{\varphi}

\begin{document}


\title{Nonlinear memory in cell division dynamics across species}

\author{Shijie Zhang}
\thanks{These two authors contributed equally.}
\affiliation{Department of Mathematics, Massachusetts Institute of Technology, 182 Memorial Dr, Cambridge, MA 02139, USA}
\author{Chenyi Fei}%
\thanks{These two authors contributed equally.}
\affiliation{Department of Mathematics, Massachusetts Institute of Technology, 182 Memorial Dr, Cambridge, MA 02139, USA}
\author{Jörn Dunkel}
 \email[Corresponding author: ]{dunkel@mit.edu}
\affiliation{Department of Mathematics, Massachusetts Institute of Technology, 182 Memorial Dr, Cambridge, MA 02139, USA}

\begin{abstract}
Regulation of cell growth and division is essential to achieve cell-size homeostasis. Recent advances in imaging technologies, such as  ``mother machines" for bacteria or yeast, have allowed long-term tracking of cell-size dynamics across many generations, and thus have brought major insights into the mechanisms underlying cell-size control. However, understanding the governing rules of cell growth and division within a quantitative dynamical-systems framework remains a major challenge. Here, we implement and apply a framework that makes it possible to infer stochastic differential equation (SDE) models with Poisson noise directly from experimentally measured time series for cellular growth and divisions.
To account for potential nonlinear memory effects, we parameterize the Poisson intensity of stochastic cell division events in terms of both the cell's current size and its ancestral history. By applying the algorithm to experimentally measured cell size trajectories, we are able to quantitatively evaluate the linear one-step memory hypothesis underlying the popular ``sizer",``adder", and ``timer" models of cell homeostasis. For \emph{Escherichia coli} and \emph{Bacillus subtilis} bacteria,  \textit{Schizosaccharomyces pombe} yeast and \emph{Dictyostelium discoideum} amoebae,  we find that in many cases the inferred stochastic models have a substantial nonlinear memory component.
This suggests a need to reevaluate and generalize the currently  prevailing  linear-memory paradigm of cell homeostasis. More broadly, the underlying inference framework is directly applicable to identify quantitative  models for stochastic jump processes in a wide range of scientific disciplines.
\end{abstract}


\maketitle

\section{Introduction}

{S}tochasticity is an intrinsic and essential feature of cellular dynamics, from gene regulation~\cite{paulsson2004summing} and biochemical reactions \cite{heinrich2012regulation,qian2011nonlinear, lord2019stochastic, yan2019kinetic} to the biomechanical control of cell size~\cite{amir2014cell,cadart2019physics,adiciptaningrum2015stochasticity,le2021quantitative}.  Over the past two decades, major technological advances in microfluidics and microscopy have enabled long-term tracking of cell growth and divisions \cite{wang2010robust,tanaka2021dynamic,nordholt2020biphasic,nakaoka2017aging}, movements \cite{rabut2004automatic,meijering2009tracking}, and gene expression \cite{bar2012studying} at high temporal resolution. The vast time-series data generated by such experiments has elevated the development of stochastic models of cellular dynamics~\cite{kar2023using,ho2018modeling,jun2018fundamental}, enabling the identification and discrimination of regulatory mechanisms that determine cell growth~\cite{micali2018dissecting,si2019mechanistic} and proliferation \cite{hallatschek2023proliferating}. A prime example is the widely considered ``sizer",``adder", and ``timer" models \cite{facchetti2017controlling,jun2015cell} of cell-size control, which have been extensively characterized in ``mother-machine" experiments~\cite{wang2010robust,duran2020slipstreaming,kaiser2018monitoring} that allow for high-throughput measurements of cell size dynamics over hundreds of generations \cite{wang2010robust,iyer2014scaling, nakaoka2017aging}.  However, despite such major progress,  there still exist fundamental open questions regarding the nonlinear and multi-generational memory \cite{elgamel2023multigenerational} effects  that are not captured by the currently prevailing standard models of cell-size homeostasis. 
\par

Here, we provide a generic framework for answering these and related questions through stochastic differential equation (SDE) inference, by implementing a Bayesian model inference scheme for stochastic jump processes \cite{schaft2000introduction}. The framework is directly applicable to cell-size trajectory data from recent mother-machine experiments~\cite{wang2010robust,duran2020slipstreaming,kaiser2018monitoring} as well as to structurally similar time-series data from other scientific disciplines (SI Appendix).  Applying this framework to growth-and-division trajectories for bacteria \cite{wang2010robust,nordholt2020biphasic}, yeast \cite{nakaoka2017aging} and amoebae \cite{tanaka2021dynamic}, we find that the conventional ``sizer",``adder", and ``timer" mechanisms cannot account for key statistical correlations in the experimentally observed cell-size dynamics. Instead, across all the species analyzed below, the  models identified by Bayesian inference consistently exhibit substantial nonlinear memory effects. Our findings suggest a need for extending   prevalent cell-division paradigms and for exploring the molecular and biophysical mechanisms underlying nonlinear memory in future experimental studies.

``Sizer",``adder", and ``timer" models have been shown to be powerful at explaining biologically relevant subsets of statistical observables \cite{jun2015cell,facchetti2017controlling}, such as the average size gains conditional on a cell's initial size. While the conceptual strength and appeal of those models lie in their simplicity,  predictive limitations arise from the way in which they implement memory~
\cite{amir2014cell,marantan2016stochastic}. ``Sizer" mechanisms posit  that a cell aims to divide consistently at a constant target size, independent of its size at birth~\cite{facchetti2017controlling}. In contrast, ``adder" models assume that a cell adds a fixed volume (half the target cell-size) during each cell cycle, so that size fluctuations arising from not perfectly symmetric divisions become  asymptotically suppressed \cite{barber2021modeling}. ``Timer" models hypothesize that cells grow for a fixed period before dividing again \cite{jun2015cell,marantan2016stochastic}.  
``Timer" mechanisms  become ineffective at maintaining size homeostasis if cell growth is exponential, and they effectively reduce to an adder mechanism if the cell growth is linear \cite{marantan2016stochastic, ovrebo2022cell}. Despite their biological differences, these three models generally fall into a broader class of models where the target division size depends linearly on the cell size at birth~
\cite{amir2014cell,marantan2016stochastic}. Thus, by construction, they cannot account for the possibility of nonlinear and multi-generational memory that may be important in cell size control.  As shown below, combining a suitably designed Bayesian inference scheme with experimental time-series data makes it possible to identify quantitative SDE models that can reveal non-trivial memory effects. 
\par
\begin{figure*}[htbp]
\centering
\includegraphics[width=.9\textwidth]{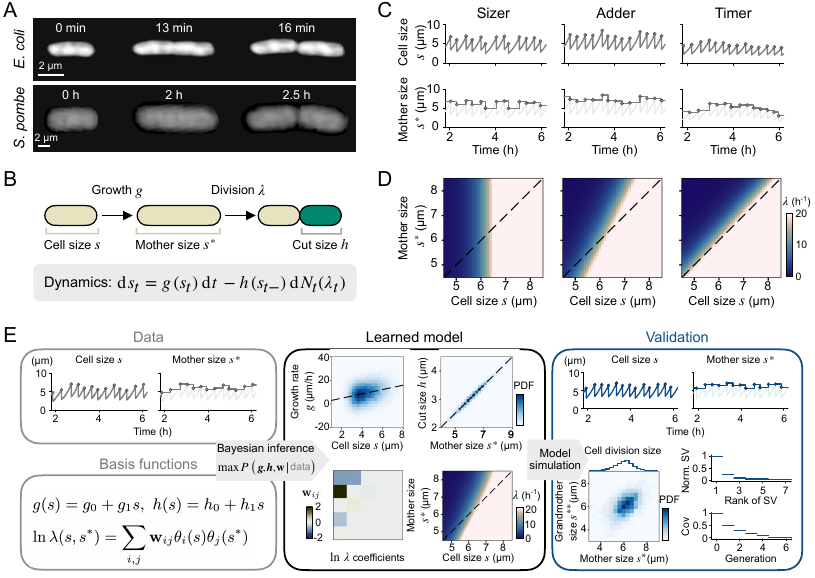}
\caption{Learning cell growth and division dynamics using a Bayesian inference framework. 
(\textit{A}) Snapshots of \textit{E.~coli} (\textit{Top}; from \cite{taheri2015cell}) and \textit{S.~pombe} (\textit{Bottom}; from \cite{nakaoka2017aging}) single cells undergoing growth and division. 
(\textit{B}) Illustration of cell growth and division dynamics described by a stochastic differential equation. Cell size $s_t$ grows deterministically with a rate of $g(s_t)$. Cell division is modeled by a Poisson process with a rate $\lambda$ that depends on current size $s$ and mother size $s^{\ast}$ denoting cell size at last division.
When division occurs, the cell is reduced by a deterministic amount of size $h(s_{t^{-}})$.
(\textit{C}) Typical trajectories of cell size $s_t$ (\textit{Top}) and mother size $s^{\ast}$ (\textit{Bottom}) for the ``sizer", ``adder", and ``timer" models simulated using Eq.~\ref{eqn:sde_main}. 
(\textit{D}) Cell division rates $\lambda$ of the corresponding models in \textit{C}. Black dashed lines indicate $s = s^{\ast}$. 
The division rate $\lambda$ of the ``sizer" model has no memory, depending only on current cell size $s_t$. For both ``adder" and ``timer" models, $\lambda$ depends linearly on the mother size $s^{\ast}$.
(\textit{E}) Inference and analysis of simulation data. \textit{Left}: Inputs are time series of cell size $s$ and mother size $s^{\ast}$. Growth rate $g(s)$ and cut size $h(s)$ are fit by linear functions. The logarithm of the division rate $\ln \lambda(s, s^{\ast})$ is decomposed using suitable spectral basis functions. \textit{Middle}: Standard linear regression is used to determine the coefficients $g_{0,1},h_{0,1}$ for $g(s)$ and $h(s)$. To avoid overfitting, a Bayesian inference algorithm with a sparsity-promoting prior determines the coefficients $w$ for $\ln \lambda(s, s^{\ast})$. \textit{Right}: The resulting model can be validated against the input data in the \textit{Left} panel.
}
\label{fig1}
\end{figure*}

Over the last few years, various data-driven  frameworks for inferring ordinary~\cite{brunton2016discovering, mangan2016inferring, mangan2017model,mangan2019model,reinbold2019data,reinbold2020using} or partial~\cite{supekar2023learning,maddu2022stability,maddu2021learning} differential equations have been developed based on least-square fitting and sparse regression techniques~\cite{maddu2021learning,mangan2016inferring,mangan2019model}. These methods have shown promise in identifying quantitative models of complex active matter  systems~\cite{supekar2023learning,maddu2022learning,golden2023physically}  exhibiting approximately deterministic dynamics. Aiming to extend these approaches to inherently noisy biological processes~\cite{paulsson2004summing, xia2021kinetic,wilkinson2018stochastic,li2017review},  recent work \cite{frishman2020learning,ronceray2024learning, bruckner2020inferring} has focused on developing inference schemes for continuous stochastic processes driven by Gaussian white noise. Applications of these techniques to biophysical data have yielded important insights into cell migration and interaction processes \cite{bruckner2021learning}. Here, we extend these efforts to discontinuous jump processes that provide an effective mathematical description of cell growth and division dynamics. Unlike neural-network based approaches to SDE inference \cite{ryder2018black, tzen2019neural, jia2019neural, kidger2021neural}, which require large training data sets that can be difficult to obtain for biological systems, we focus here on a Bayesian inference approach that work efficiently on recently reported time-series measurements for bacteria \cite{wang2010robust,nordholt2020biphasic}, yeast \cite{nakaoka2017aging} and amoebae \cite{tanaka2021dynamic}. In short,  our model inference and selection framework takes cell-size trajectories as input and gives  the governing equations as output (Fig.~\ref{fig1}). By integrating sparse Bayesian inference with orthogonal basis-function representations, we identify SDE models with inhomogeneous Poisson noise that agree with the experimentally observed growth and division dynamics and cell-size correlations.
The inferred models reveal a spectrum of cell-division strategies, highlighting  nonlinear memory of the mother-cell size in various organisms. 
\par
More generally, due to its system-agnostic formulation, the inference framework provided here is broadly applicable beyond biological data. As practical guidance for future applications, we provide several example demonstrations, including applications to earthquake and internet time-series data, in the SI Appendix.

\section{Results}

\subsection{An SDE description of cell growth and division dynamics \label{SDE_description}}
Prokaryotic and eukaryotic cells come in various sizes and shapes, but they typically maintain size homeostasis by coordinating growth and division \cite{amir2014cell,taheri2015cell}. During the growth phase, cells replicate DNA and synthesize proteins to increase biomass \cite{cooper2012bacterial}. Once they have completed DNA replication and have accumulated enough proteins and other important resources, the mother cells divide and allocate resources into two (or possibly more) daughter cells (Fig.~\ref{fig1}A), which then repeat this cycle to proliferate.
To describe cell-size dynamics across multiple generations for cells undergoing binary division, we consider a stochastic differential equation with an inhomogeneous Poisson noise (Fig.~\ref{fig1}B)
\begin{equation}
\text{d} s_{t} = g(s_{t}) \text{d} t \, - h(s_{t^{-}}) \text{d} \mathcal{N}_t(\lambda_t), \label{eqn:sde_main}
\end{equation}
where $s_t$ denotes cell size at time $t$. The first term on the right-hand-side describes deterministic cell growth with a rate function $g(s_{t})$. 
The Poisson counting process $\mathcal{N}_t$ in Eq.~(\ref{eqn:sde_main}) is parameterized by a history-dependent rate function $\lambda_t$ and describes the occurrence of cell-division events at time $t$, where $\text{d} \mathcal{N}_t = 1$ indicates a division event and $\text{d} \mathcal{N}_t = 0$ otherwise.
Consequently, when cell division occurs, the size of the mother cell right before division, $s_{t^{-}}$, is reduced by a cut size $h(s_{t^{-}})$, reflecting the born size of the untracked daughter cell.
For example, symmetric cell division corresponds to $h(s_{t^{-}}) = s_{t^{-}} / 2$.
Depending on the functional forms of $g$, $h$ and $\lambda$, Eq.~(\ref{eqn:sde_main}) predicts typical cell-size trajectories  with continuous growth interrupted by   discontinuous (negative) ``jumps" due to division (Fig.~\ref{fig1}C).
\par
Within the modeling framework defined by Eq.~(\ref{eqn:sde_main}), the inhomogeneous Poisson intensity $\lambda_t$ characterizes and distinguishes between different cell division strategies. In particular, larger values of  $\lambda_t$ increase the likelihood of division at time $t$. 
Since each cell receives a set of  biochemicals, including DNA and proteins, from its mother cell \cite{aguilaniu2003asymmetric,spokoini2012confinement}, this inheritance of cellular assets could in principle allow for a form of ``cell division memory" where the target division size of a cell depends on its ancestral history. 
To capture such cellular memory, we define a set of new time series $s_t^{\ast \cdots \ast}$ that track the cell sizes at previous  divisions. 
We designate $s_t^{\ast}$ as the mother size (Fig.~\ref{fig1}C) and $s_t^{\ast \ast}$ as the grandmother size, with each additional superscript $*$ indicating an additional prior  generation in the family tree. 
Below we focus on a minimal model of cell-division strategy with one-generation memory, where $\lambda_t = \lambda(s_t, s_t^{\ast})$ is determined by both the current cell size $s_t$ and the mother size $s_{t}^{\ast}$. Results for inferring models with multigenerational memory are provided in the SI Appendix.
Notably, the prevailing models of ``sizer", ``adder", and ``timer" can all be described by a special case of the one-generation memory model $\lambda(s_t,s_{t}^{\ast}) = H(s_t-\Tilde{s}(s_{t}^{\ast}))$ (Fig.~\ref{fig1}C--D), where $H$ is a step-like function that transits from $H=0$ to $H\to \infty$ as the cell size $s_t$ exceeds the target division size $\Tilde{s}(s_{t}^{\ast})=c\,s_t^{\ast}+\Delta$. We will refer to this as the linear-memory model below. Specifically, the ``sizer" corresponds to $c=0, \Delta > 0$, the ``adder" to $c=1/2, \Delta > 0$, and the ``timer" (with exponential growth) to $c=1, \Delta = 0$.
In general, our minimal model $\lambda_t = \lambda(s_t, s_t^{\ast})$ is capable of describing more complex cell-division strategies, such as those with a nonlinear $\tilde{s}(s_t^\ast)$, beyond these three conventional paradigms. Moreover, our approach readily extends to modeling multi-generational memory by incorporating higher-order $s_t^{\ast \cdots \ast}$ (SI Appendix, section \ref{sec:S1}). We next describe how one can infer an SDE models as in Eq.~(\ref{eqn:sde_main}) from experimentally measured time series~$s_t$.

\subsection{Data-driven discovery of SDE models}
Individual cells in many species grow linearly or exponentially \cite{kar2021distinguishing,cooper2006distinguishing} but more general growth models have also been proposed \cite{vuaridel2020computational}. To allow for linear, exponential and nonlinear growth dynamics, our inference framework assumes a generic quadratic growth rate function $g(s_{t}) = g_0 + g_1 s_{t} + g_2 s_{t}^2$ in Eq.~(\ref{eqn:sde_main}).  Specifically, $g_1=g_2=0$ corresponds to linear growth and $g_0=g_2=0$ to exponential growth.
Our parameter inference for experimental data from \textit{E.~coli} \cite{wang2010robust}, \textit{B.~subtilis} \cite{nordholt2020biphasic}, \textit{S.~pombe} \cite{nakaoka2017aging}, \textit{D.~discoideum} \cite{tanaka2021dynamic} showed that the quadratic term is negligible for all these species ({SI Appendix}, Fig.~\ref{SIfig_g_coefs}). We therefore restrict the  main text discussion to linear rate functions $g(s_{t}) = g_0 + g_1 s_{t}$ from now on. Similarly,  the cut size of a cell at division is also modeled by a linear function $h(s_{t^{-}}) = h_0 + h_1 s_{t^{-}}$.  Given a cell-size time series~$s_t$, it is straightforward to estimate   the coefficients $g_i$ and $h_i~(i=0,1)$ by applying  linear regression to the continuous growth phase and the discontinuous jumps, respectively (Fig.~\ref{fig1}E and {SI Appendix}, section \ref{sec:S1}). It thus remains to identify and constrain a division rate model for the Poisson counting  process $\mathcal{N}_t$.
\par
To infer the inhomogeneous cell-division rate $\lambda(s_t,s_{t}^{\ast})$ that determines the statistics of $\mathcal{N}_t$ in Eq.~\ref{eqn:sde_main}, we combine basis-function representation and Bayesian inference. The resulting  computational framework  takes time-series data of cell sizes $s_t$ and $s_t^{\ast}$ as input and determines an expression for $\lambda(s_t,s_{t}^{\ast})$ as output. To ensure that $\lambda(s_t,s_{t}^{\ast})$ is always positive, we work with $\ln \lambda$ and represent it as  
\begin{equation}
\label{eqn:spectralrepresentation}
\ln \lambda( s_{t}, s_{t}^{\ast}) = 
\sum_{i,j} \mathbf{w}_{ij}\theta_i(s_t)\theta_j(s_t^{\ast}) 
\end{equation}
where $\{\theta_i\}$ are orthogonal polynomials that we constructed from data using the modified Gram-Schmidt procedure \cite{bjorck1994numerics} (see {SI Appendix}, section \ref{sec:S1} and Fig.~\ref{SIfig_orthobasis}). The mode coefficients~$\mathbf{w}_{ij}$  encode the information about cell-division strategies. Generally, our approach is insensitive to the specific choices of basis functions provided that they are compatible with the data ({SI Appendix}, Fig.~\ref{SIfig_varbasis}). Using the representation in Eq.~(\ref{eqn:spectralrepresentation}), we next seek to find the most probable coefficients $\hat{\mathbf{w}}$ given the observed data $s_t$, or in the language of Bayesian statistics, maximize the posterior probability $P(\mathbf{w} | s_t)$. According to Bayes' theorem, we decompose the posterior $P(\mathbf{w} | s_t)\propto P(s_t | \mathbf{w}) P(\mathbf{w})$ into two components: a likelihood function $P(s_t | \mathbf{w})$ for the observed data based on the model coefficients $\mathbf{w}$, and a prior probability $P(\mathbf{w})$ that encodes our preference for the desired coefficients $\mathbf{w}$.
Since we assume that the division events are generated by an inhomogeneous Poisson process, we derive the likelihood function to be
\begin{equation}
    P(s_t | \textbf{w}) = \exp(-\int \lambda(s_{t}, s_{t}^{\ast}) dt)\prod_{i} \lambda(s_{t}, s_{t}^{\ast})|_{t=\tau_{i}}, \label{eqn:likelihood} 
\end{equation}
where the first exponential term accounts for the probability of the non-dividing growth phase, and the second multiplication term represents the probability density of cell division occurring at times $\{\tau_i\}$. 

To prevent overfitting to noisy data, we impose sparsity on~$\mathbf{w}$, favoring smooth functions  $\lambda(s_{t}, s_{t}^{\ast})$ with a small number of modes. 
Following previous works on sparse Bayesian inference \cite{pearl1988probabilistic,cowell2007probabilistic,babacan2012sparse}, we employ a Gaussian prior 
\begin{equation}
P(\mathbf{w}) \propto \prod_{i,j} \exp( - \frac{\mathbf{w}_{ij}^2}{2\boldsymbol{\sigma}_{ij}^2} ), \label{eqn:gaussian_prior}
\end{equation}
although our approach is robust against alternative choices of sparsity-promoting priors ({SI Appendix}, Fig.~\ref{SIfig_varreg}).
The variance hyperparameters $\boldsymbol{\sigma}_{ij}^2$ in Eq.~(\ref{eqn:gaussian_prior}) control the level of sparsity.  
Specifically, small values of $\boldsymbol{\sigma}_{ij}^2$ encourage the associated coefficients $\mathbf{w}_{ij}$ to stay close to zero, thereby facilitating the truncation of irrelevant modes in Eq.~(\ref{eqn:spectralrepresentation}).
However, the values of $\boldsymbol{\sigma}$ are unknown {\it a priori}.
Thus, we treat $\boldsymbol{\sigma}$ as hidden variables and exploit the Expectation-Maximization algorithm to iteratively update our estimation of $\boldsymbol{\sigma}$ and compute the most probable coefficients $\hat{\mathbf{w}}$ \cite{moon1996expectation,do2008expectation}. After learning $\hat{\mathbf{w}}$, we can reconstruct the cell-division strategy $\lambda$ using Eq.~(\ref{eqn:spectralrepresentation}) (Fig.~\ref{fig1}E).

\subsection{Model selection with information criteria \label{Model_selection}}
Drawing from previously proposed frameworks on sparse identification of dynamical systems \cite{brunton2016discovering,quade2018sparse}, we apply sequential thresholding on the coefficients $\hat{\mathbf{w}}$ to produce a series of models with decreasing complexities in linear time. 
Although a model with higher complexity, or more terms in the expansion in Eq.~(\ref{eqn:spectralrepresentation}), generally provides a better fit to data, it is also more susceptible to overfitting, which can compromise a model's predictive power and generalization performance~\cite{hastie2009elements}. This trade-off is also an important notion in machine learning \cite{srivastava2014dropout,mehta2019high,bartlett2020benign,murphy2022probabilistic}. Thus, the goal is to select a parsimonious model that best explain the data with the least number of terms. One advantage of Bayesian inference is that it comes naturally with a principled framework, known as the Bayesian information criterion (BIC), for model selection \cite{neath2012bayesian, weakliem1999critique}.
Building on conventional BIC, we employ a modified (negative) BIC score $ \sim  \ln P(s_t | \hat{\mathbf{w}} ) - 1/2\ln |\hat{\mathbf{H}}|$ for stochastic models with non-Gaussian noise ({SI Appendix}, section \ref{sec:S1}). Here, $\hat{\mathbf{H}}$ denotes the Hessian of the log posterior evaluated at $\hat{\mathbf{w}}$ and its log-determinant $\ln |\hat{\mathbf{H}}|$ serves as a penalty term for the number of model parameters. A learned model with the highest modified BIC score best balances model accuracy and complexity.

\subsection{Validation on synthetic data}
\begin{figure*}[htbp]
\centering
\includegraphics[width=0.9\textwidth]{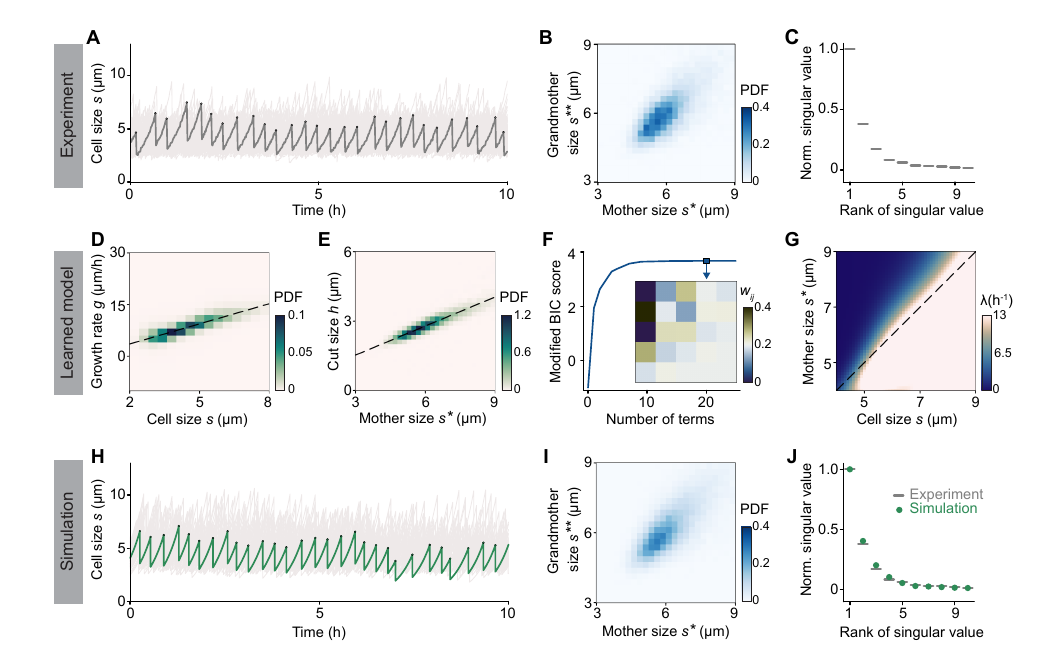}
\caption{Growth-and-division model learned from \textit{E.~coli} experimental data\cite{wang2010robust} reveals a strong nonlinear memory of mother cell size in division. 
(\textit{A}) Trajectories of cell sizes from 265 independent \textit{E.~coli} experiments are shown in light grey lines and a
typical trajectory is highlighted. The doubling time is approximately 20~min. 
(\textit{B}) Joint probability distribution of mother size $s^{\ast}$ (cell size at last division) and grandmother size $s^{\ast \ast}$ (cell size at the division before last division) shows a strong correlation between two consecutive division sizes. 
(\textit{C}) The 10 largest singular values of the joint probability distribution in \textit{B}. There is more than one singular value significantly greater than 0, indicating that dividing cells have substantial memory of their mother sizes. 
(\textit{D}) Heatmap of cell size $s$ versus growth rate $g$ from input data. The dashed line shows the learned linear growth rate $g(s)$, corresponding to exponential growth in \textit{E.~coli}. 
(\textit{E}) Heatmap of mother size $s^{\ast}$ versus cut size $h$ from input data. The dashed line shows the learned linear cut size $h(s)$, which is approximately symmetric division for \textit{E.~coli}.
(\textit{F}) Modified Bayesian information criterion (BIC) scores (see {SI Appendix}, section \ref{sec:S1} for details) of models with different parameter complexity (number of terms). The square marker indicates the selected model with the highest BIC score, whose coefficients $\mathbf{w}$ are shown in the inset.
(\textit{G}) The learned division rate $\lambda$ corresponding to the selected model in \textit{F} demonstrates nonlinear memory effects. Black dashed line indicates $s = s^{\ast}$.
(\textit{H}) -- (\textit{J}) Simulation results of the selected learned model of \textit{D} -- \textit{G}. Plots correspond to \textit{A} -- \textit{C}, respectively. 
}
\label{fig2}
\end{figure*}

We first demonstrate our SDE model inference and selection pipeline on synthetic cell-size trajectory data.
To mimic the dynamics of cell growth and division, we simulate the SDE model Eq.~(\ref{eqn:sde_main}) with exponential growth 
and symmetric division, which are commonly observed in living cells \cite{campos2014constant, taheri2015cell,deforet2015cell,stewart2005aging}.
We choose a sigmoid function for the cell division rate $\lambda(s_t,s_{t}^{\ast}) = \lambda_{\max}[1+\tanh\beta(s_t-\Tilde{s}(s_{t}^{\ast}))]/2$,
which describes the growth mode $\lambda\approx0$ when the cell size is much smaller than the target division size $\Tilde{s}$, that is $\beta(s_t- \Tilde{s}) \ll 0$, and approaches the maximal rate of division $\lambda \approx \lambda_{\max}$ when $\beta(s_t- \Tilde{s}) \gg 0$. As discussed above, we use a linear function $\Tilde{s}(s_{t}^{\ast}) = s_{t}^{\ast}/2+\Delta$ balancing the mother size and the average division size to simulate a model of ``adder" ({Section.~\ref{SDE_description}}). The ``adder" principle has been shown to be a common size-control strategy shared by cells from all kingdoms of life, ranging from archaeal cells \cite{eun2018archaeal}, to bacterial cells \cite{wang2010robust,taheri2015cell,nordholt2020biphasic}, to eukaryotic amoeba and yeast cells \cite{soifer2016single,tanaka2021dynamic}, to mammalian cells \cite{varsano2017probing,cadart2018size}.
Indeed, simulations of the ``adder" model yield cell-size trajectories that closely resemble the experimental data, with moderate stochasticity in the size at division (Fig.~\ref{fig1} C--D and Fig.~\ref{fig2}A).

We take the simulated trajectories as input data (Fig.~\ref{fig1}E \textit{Left}) to learn the growth rate $g(s_t)$, cut size $h(s_{t^{-}})$ and division rate $\lambda(s_t,s_{t}^{\ast})$ of the SDE model following the framework above (Fig.~\ref{fig1}E \textit{Middle}). For functions $g(s_t)$ and $h(s_{t^{-}})$, linear regression faithfully reproduce the ground-truth exponential growth and symmetric division, respectively. The learned division rate $\lambda(s_t,s_{t}^{\ast})$ has a sparse representation with only a few nonzero coefficients and agrees  quantitatively with the ground-truth sigmoid function.

To further validate the inferred model, we re-simulate Eq.~(\ref{eqn:sde_main}) with the learned $g$, $h$, and $\lambda$ to generate new cell-size trajectories, and we compare their statistics  with those of input data (Fig.~\ref{fig1}E \textit{Right}). Specifically, we examine the joint distribution $P(s^{\ast}, s^{\ast \ast})$ of mother size $s^{\ast}$ and grandmother size $s^{\ast\ast}$ in the re-simulated trajectories, finding a close match with the input data ({SI Appendix}, Fig.~\ref{SIfig_simulateddata}). As expected, both distributions show a positive correlation between the sizes of mother and grandmother cells, representing the cell-division memory in the ground-truth model.
To further quantify the cell-size memory in division, we compute the singular value spectrum of the joint distribution, which illustrates the number of independent ``modes" needed to reconstruct $P(s^{\ast}, s^{\ast \ast})$ and their relative importance. If there is no memory, meaning that the cell sizes at two consecutive divisions are independent, then $P(s^{\ast}, s^{\ast \ast}) = P(s^{\ast})P(s^{\ast \ast})$, and thus there is only one nonzero singular value. Indeed, the singular value spectra of both the input and the re-simulated distributions show a characteristic power law decay with more than one singular value significantly larger than zero ({SI Appendix}, section \ref{sec:S2}), indicating the presence of memory in the cell-size division process. The close agreement between the input data and the model simulations demonstrates that our inference framework is effective in learning a SDE description of time-series data that contain both continuous and discrete-time dynamics. We next apply this framework to  recent experimental data to quantify and classify the division strategies of different organisms.

\subsection{Application to mother-machine data for bacteria and yeast} 
\begin{figure*}[htbp]
\centering
\includegraphics[width=0.9\textwidth]{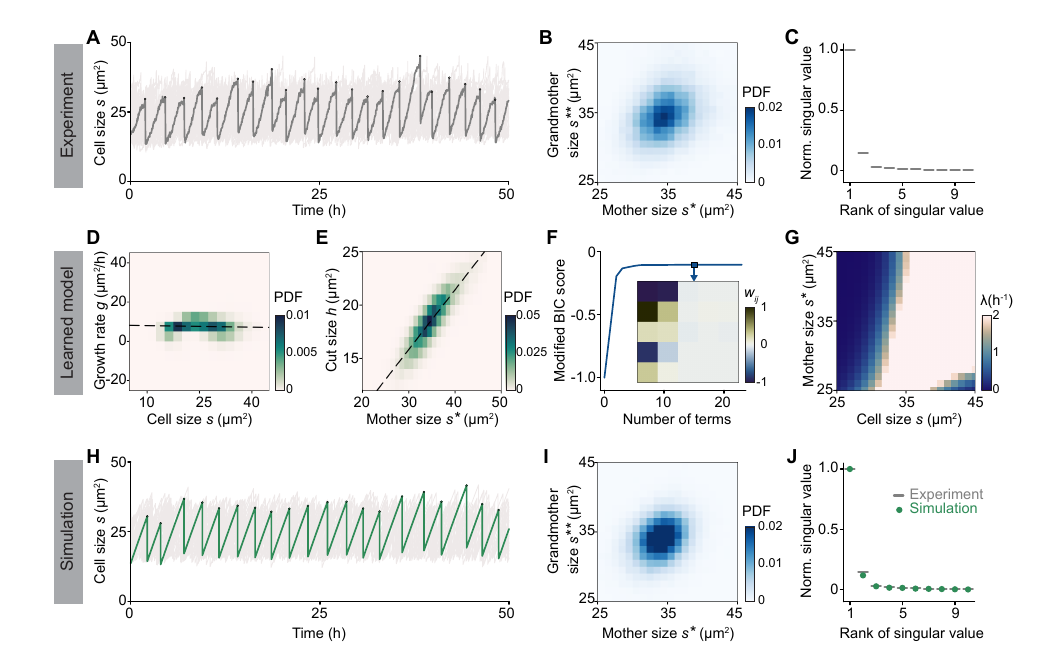}
\caption{Growth-and-division model learned from \textit{S.~pombe} experimental data\cite{nakaoka2017aging} shows that cell division has a weak memory of mother cell size. 
(\textit{A}) Trajectories of cell sizes from 50 independent \textit{S.~pombe} experiments are shown in light grey lines (1604 experiments in total) and a
typical trajectory is highlighted. The doubling time is approximately 2~h. 
(\textit{B}) Joint probability distribution of mother size $s^{\ast}$ and grandmother size $s^{\ast \ast}$ shows a weak correlation between two consecutive division sizes. 
(\textit{C}) The 10 largest singular values of the joint probability distribution in \textit{B}. There are only two singular values significantly greater than 0, indicating that dividing cells have weak memory of their mother sizes. 
(\textit{D}) Heatmap of cell size $s$ versus growth rate $g$ from input data. The dashed line shows the learned linear growth rate $g(s)$, corresponding to linear growth. 
(\textit{E}) Heatmap of mother size $s^{\ast}$ versus cut size $h$ from input data. The dashed line shows the learned linear cut size $h(s)$, which is approximately symmetric division for \textit{S.~pombe}.
(\textit{F}) Modified BIC scores of models with different parameter complexity. The square marker indicates the selected model with the highest BIC score, whose coefficients $\mathbf{w}$ are shown in the inset.
(\textit{G}) The learned division rate $\lambda$ corresponding to the selected model in \textit{F} demonstrates that \textit{S.~pombe} has weak memory of its mother size, similar to an adder-sizer mixture. 
(\textit{H}) -- (\textit{J}) Simulation results of the selected learned model of \textit{D} -- \textit{G}. Plots correspond to \textit{A} -- \textit{C}, respectively. 
}
\label{fig3}
\end{figure*}

Cell-size trajectories can be accurately measured using mother machines~\cite{spivey2017aging, spivey20143d,nobs2014long,zhang2012single}. These high-throughput microfluidic platforms can track the lineage of old-pole cells over hundreds of generations by trapping founder cells in one ended growth channels and washing progeny cells away at the open ends \cite{spivey2017aging, spivey20143d,nobs2014long,zhang2012single}. 
Here, we demonstrate the broad applicability of our inference pipeline on time-series data obtained from previous mother-machine experiments conducted with the bacterium \textit{Escherichia coli} (Fig.~\ref{fig2}) \cite{wang2010robust} and the fission yeast \textit{Schizosaccharomyces pombe} (Fig.~\ref{fig3}) \cite{nakaoka2017aging}. 
\par
The raw data from these experiments can be used for SDE inference after performing a few elementary pre-processing steps ({SI Appendix}, section \ref{sec:S1} and Fig.~\ref{SIfig_preprocessing}). First, we filter out trajectory segments that do not show periodic growth and division. Furthermore, to ensure consistency in the inference process and to prevent incorporating data from multiple physiological states, we also discard segments of trajectories that show a transient chaining phenotype for which cells grow to an abnormally large size before division \cite{norman2015stochastic}.  The resulting input time-series~$s_t$ can be used to identify cell-size ``jumps"  division events, also yielding the time series of mother sizes $s_t^\ast$ for learning cell-division memory. Representative examples of trajectories $s_t$ used for further inference analysis are shown in  Fig.~\ref{fig2}A and \ref{fig3}A.
\par
Next, we directly apply the inference framework to the pre-processed trajectory data $s_t$ to learn SDE models in the form of Eq.~(\ref{eqn:sde_main}) that describe the experimental cell growth and division dynamics. 
The inferred growth rate functions $g(s_t)$ show that \textit{E.~coli} follows exponential growth (Fig.~\ref{fig2}D) whereas \textit{S.~pombe} undergoes linear growth (Fig.~\ref{fig2}D). Both \textit{E.~coli} and \textit{S.~pombe} exhibit symmetric cell division as suggested by the inferred cut size $h(s_{t^-})\approx s_{t^-}/2$ (Fig.~\ref{fig2}E and \ref{fig3}E).  Our inference framework identifies  a set of candidate models for the cell division rate $\lambda(s_t,s_{t}^{\ast})$. These candidate models differ in their parameter complexity, and we select an optimal model based on BIC to avoid overfitting (Fig.~\ref{fig2}F and \ref{fig3}F, see also {Section.~\ref{Model_selection}}).

\subsection{SDE inference reveals nonlinear cell-division memory} 
For \textit{E.~coli}, the inferred cell division rate $\lambda(s_t,s_{t}^{\ast})$ shows a highly nonlinear memory of the mother size $s_{t}^{\ast}$ (Fig.~\ref{fig2}G), indicating that cells smaller than the population average converge more rapidly to the average size compared to those larger than the population average.
For \textit{S.~pombe}, the learned $\lambda(s_t,s_{t}^{\ast})$ shows a weak dependence on the mother size $s_{t}^{\ast}$ and closely resemble a sizer-like linear-memory model, consistent with previous studies \cite{jun2015cell, pombe1977control}.
\par
To further validate the inferred SDE models, we generate cell-size trajectories by simulating our inferred models (Fig.~\ref{fig2}H and \ref{fig3}H). The simulated data quantitatively captures the distributions of division sizes and generation times observed in each experimental dataset (Fig.~\ref{fig2}I and \ref{fig3}I, {SI Appendix}, Fig.~\ref{SIfig_amoeba}--\ref{SIfig_statistics}). To assess the degree of memory in cell division, we perform singular-value decomposition of the joint distributions of two consecutive division sizes. The resulting singular-value spectra agrees well between the experimental and simulation data (Fig.~\ref{fig2}J and \ref{fig3}J), confirming a stronger cell-division memory in \textit{E.~coli} than in \textit{S.~pombe}. Specifically,  for \textit{E.~coli} cell sizes at two consecutive divisions are highly correlated,  whereas  consecutive divisions in \textit{S.~pombe} are nearly independent statistically.

\subsection{One- vs. multi-generation memory}
The pronounced cell-division memory in \textit{E.~coli} raises the question whether the one-generation memory model is sufficient and necessary to fully capture the experimental data. To explore this question, we compare inference results among three models with different generations of memory: one with no-memory $\lambda_t = \lambda(s_t)$, one with one-generation memory~$\lambda_t=\lambda(s_t, s_t^{\ast})$, and one with two-generation memory $\lambda_t = \lambda(s_t, s_t^{\ast}, s_t^{\ast\ast})$, where the current size $s_t$, the mother size $s_t^\ast$, and the grandmother size $s_t^{\ast\ast}$ represent progressively higher orders of memory in the cell-division rate~$\lambda$ ({SI Appendix}, Fig.~\ref{SIfig_012gen_memory}). 
As expected, the no-memory model fits poorly to the \textit{E.~coli} data, indicated by a low BIC score, as this minimal model fails to capture the correlation of cell sizes at consecutive divisions.
Adding one-generation memory of the mother size $s_t^\ast$ significantly improves the fitting, leading to a learned model capable of generating time series of cell sizes that are almost statistically identical to the experimental data (Fig.~\ref{fig2}). Perhaps surprisingly, adding memory of the grandmother size $s_t^{\ast\ast}$ does not further improve the BIC score of the model, and the learned cell-division rate $\lambda(s_t, s_t^{\ast}, s_t^{\ast\ast})$ shows minor dependence on $s_t^{\ast\ast}$ ({SI Appendix}, Fig.~\ref{SIfig_012gen_memory}). Thus, our results suggest that a parsimonious model with one-generation memory effectively accounts for the observed dynamics in \textit{E.~coli}.

\begin{figure}[!htbp]
\centering
\includegraphics[width=0.5\textwidth]{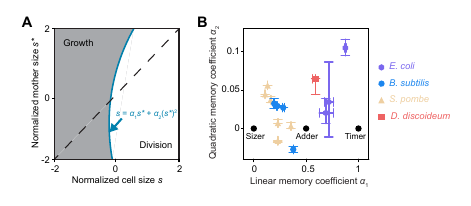}
\caption{Nonlinear memory of cell division found across species. 
(\textit{A}) Schematics of the division rate $\Tilde{\lambda}(s_t,s_t^{\ast})$ showing a distinct separation between larger values (division phase) and smaller values (growth phase). The boundary line is fitted by a quadratic function $s = \alpha_1 s^{\ast} + \alpha_2 (s^{\ast})^2$.
(\textit{B}) Plotting linear ($\alpha_1$) vs. quadratic ($\alpha_2$) fit coefficients shows that conventional linear-memory models (filled black circles) are inconsistent with experimental data, highlighting the presence of nonlinear cell-division memory  across different species (\textit{E.~coli} \cite{wang2010robust}, \textit{B.~bacillus} \cite{nordholt2020biphasic}, \textit{S.~pombe} \cite{nakaoka2017aging}, \textit{D.~discoideum} \cite{tanaka2021dynamic}). 
}
\label{fig4}
\end{figure}

\subsection{Nonlinear memory across different species}
Previous studies have used a class of linear-memory models to study and classify cell-division strategies in various organisms ({Section.~\ref{SDE_description}}) \cite{eun2018archaeal,taheri2015cell,nordholt2020biphasic,soifer2016single,tanaka2021dynamic,varsano2017probing,cadart2018size}. However, our above analysis  suggests that the cell-division rate can depend nonlinearly on the mother size, which motivates us to find a way to quantify the nonlinearity of cell memory in division.
Since the inferred division rate $\lambda(s,s^\ast)$ for both \textit{E.~coli} (Fig.~\ref{fig2}G) and \textit{S.~pombe} (Fig.~\ref{fig3}G) show a sharp transition between the growth phase (low $\lambda$) and division (high $\lambda$) phase, the information about $\lambda$ is almost fully encoded in the boundary curve $s(s^\ast)$ delineating this transition (Fig.~\ref{fig4}A). 
We extract $s(s^\ast)$ as a contour curve of $\lambda(s, s^\ast)$ such that $s(s^\ast)$ intersects with $s=s^\ast$ at the average division size $s^\ast=\langle s^\ast \rangle$ of the input data.
Subsequently, we mean-center and normalize $s(s^\ast)$, and fit a quadratic curve $s(s^\ast)\sim \alpha_1 s^\ast + \alpha_2 (s^\ast)^2$ locally around $s^\ast=\langle s^\ast \rangle$. This two-dimensional representation $(\alpha_1, \alpha_2)$ further reduces the risk of overfitting and allows us to compare and classify cell-division strategies across species.
Within this framework, the linear-memory models correspond to $\alpha_2 = 0$, with the ``sizer", ``adder", ``timer" models described by $\alpha_1=0$, $\alpha_1=1/2$, and $\alpha_1=1$, respectively (black filled circles in Fig.~\ref{fig4}B).
We apply our Bayesian inference framework to  datasets from four organisms, \textit{Escherichia coli}, \textit{Bacillus subtilis}, \textit{Dictyostelium discoideum}, and \textit{Schizosaccharomyces pombe}, and reduce the inferred cell-division rates $\lambda$ to the $(\alpha_1,\alpha_2)$ pairs. Our results reveal that \textit{E.~coli} and \textit{D.~discoideum} resemble timer-like adders with nonlinear memory of mother size, whereas \textit{B. subtilis} and \textit{S.~pombe} appear to follow a mixed sizer-adder dynamics with  weakly nonlinear memory (Fig.~\ref{fig4}B). This  example analysis for four different species illustrates how the SDE-based approach introduced makes it  possible to go beyond the extensively studied linear-memory models by introducing a new dimension to describe nonlinear memory in cell division dynamics. Due to its generic formulation, the underlying inference  framework provides a theoretical and practical foundation for quantifying the effects of mutations or environmental conditions on growth and division  memory.

\section*{Discussion \& Conclusions}
By combining orthogonal basis function representations with sparse Bayesian inference, the above inference framework can be used to learn parsimonious SDE models from stochastic time-series data exhibiting discontinuous jumps. Here, we mainly focused on applying our framework to experimentally measured  trajectories of cell sizes to discover the dynamical equations governing cell growth and division. By modeling cell division as an inhomogeneous Poisson process, our approach identifies SDE models that accurately reproduce the measured statistics of cell division sizes and generation times ({SI Appendix}, Fig.~\ref{SIfig_statistics}). The inferred SDE models generalize prevalent linear-memory models, and reveal the presence of substantial nonlinear memory in cell division dynamics across different species  (Fig.~\ref{fig4}).

\par

The above analysis demonstrates the practical potential of SDE-based model inference for  comparing cell dynamics at the species level.  Importantly,  the underlying framework can be readily extended to study cell-cell variability within the same species~\cite{susman2018individuality} and to help identify the biological mechanisms~\cite{si2019mechanistic} underlying nonlinear growth and division memory.  Previous research has identified the essential genes and proteins involved in cell division \cite{groisman2000cpeb, rothfield1999bacterial}, and there is a rapidly growing body of data on gene expression and protein concentration dynamics measured using mother machine and other microscopic devices \cite{susman2018individuality,bakshi2021tracking,sassi2022protein}. These resources can be leveraged in conjunction with Bayesian inference schemes as developed here to identify increasingly more accurate models of cellular dynamics. As a specific next step,  building on the methodology developed here, we plan to integrate cell-size data with measurements of protein content and other cellular variables in the same cells to infer multivariate models that encode the interactions between these cellular variables. Such multivariate SDE models  promise quantitative  insights into the coordination of cell division, DNA replication, and other intracellular activities.
\par
From a more general computational perspective, the broad appeal and potential of Bayesian SDE-inference schemes  as considered here lies in the fact that they can be readily modified and adapted. Our above framework for stochastic jump processes can be seamlessly interfaced with other methods of function approximation, numerical optimization, and model selection, including those extensively used in modern-day machine learning \cite{murphy2022probabilistic, mehta2019high}. For example, one can replace the spectral basis-function representation with neural networks and expectation-maximization with stochastic gradient descent, while maintaining the ability to effectively learn models from data. Such flexibility opens a wide range of applications across scientific disciplines. To illustrate the practical potential  explicitly, we present in the {SI Appendix} several example applications that show how the  framework implemented here can also be used to learn SDE representations  from a broad spectrum of real-world datasets, including online user activities, clinical visits, and earthquake records ({SI Appendix}, Fig.~\ref{SIfig_stackexchange}--\ref{SIfig_earthquake}). We therefore expect that the underlying methodology can serve as a general foundation for quantitative  model discovery in various fields, including biology, healthcare, and geoscience.

\begin{acknowledgments}
We thank Alasdair Hastewell and George Stepaniants for helpful discussions and comments.
This work is supported by the National Science Foundation DMR/MPS-2214021 (J.D.), 
Sloan Foundation Grant G-2021-16758 (J.D.), and the MathWorks Proessorship Fund (J.D.). This research received support through Schmidt Sciences, LLC (Schmidt Science Polymath Award to J.D.).
\end{acknowledgments}

\clearpage
\bibliography{ref}

\clearpage
\onecolumngrid
\begin{center}
\textbf{\large SI Appendix}
\end{center}
\setcounter{equation}{0}
\setcounter{figure}{0}
\setcounter{section}{0}
\renewcommand{\theequation}{S\arabic{equation}}
\renewcommand{\thefigure}{S\arabic{figure}} 
\renewcommand{\thesection}{\arabic{section}}
\renewcommand{\thesubsection}{\Alph{subsection}}

\newcommand{\td}[1]{\tilde{#1}}

\newcommand{\D}{\mathrm{d}}
\newcommand{\cs}[2]{
  s%
  \ifx#1\empty
  \else
    _{#1} 
  \fi
  \ifx#2\empty
  \else
    ^{#2} 
  \fi
}
\newcommand{\dCs}{\Delta S^*}
\newcommand{\Cs}{S_0}
\newcommand{\St}{s_{t}} 
\newcommand{\Stminus}{s_{t^{-}}}
\newcommand{\Sd}{s^{(-1)}}
\newcommand{\Sdd}{s^{(-2)}}
\newcommand{\Sdn}{s^{(-n)}}
\newcommand{\um}{\mathrm{\upmu m}}

\section{Sparse inference of cell growth and division dynamics \label{sec:S1}}

To describe the dynamics of cell growth and division, we employ the following stochastic differential equation (SDE) to model the temporal evolution of cell size $\cs{t}{}$:
\begin{equation}
\label{eqn:SDE}
\D \cs{t}{} = g(\cs{t}{}) \D t \, - h(\cs{t^{-}}{}) \D \mathcal{N}_t(\lambda),
\end{equation}
where $g(\cs{t}{})$ is the rate of deterministic growth, $h(\cs{t^{-}}{})$ is the change in cell size at division, and the discrete cell division events are described by an inhomogeneous Poisson counting process $\mathcal{N}_t(\lambda)$ with a time-varying intensity $\lambda$. During cell division, daughter cells can inherit certain cellular properties from their mother, which can be then maintained for several generations \cite{vashistha2021non}. To capture the potential memory in cell division, we consider a cell division rate $\lambda(\cs{t}{}, \cs{t}{\ast}, \cs{t}{{\ast\ast}}, ..., \cs{t}{\ast\dots \ast})$ that depends not only on the current cell size $\cs{}{}$ but also on the cell size $\cs{}{\ast\dots \ast}$ at previous cell divisions. Here, $\cs{t}{\ast}$ denotes the mother size, $\cs{t}{{\ast\ast}}$ denotes the grandmother size, and each additional $*$ in the superscript indicates one more generation in the family tree. 

\subsection{Sparse Bayesian inference of intensity $\lambda$}
Without loss of generality, we will assume that $\lambda$ contains only one generation of memory in this section, taking the form of $\lambda(\cs{t}{}, \cs{t}{\ast})$. The procedures outlined below can be readily extended to include multiple generations of memory (Fig.~\ref{SIfig_012gen_memory}). To infer the time-varying cell division rate $\lambda(\cs{t}{}, \cs{t}{\ast})$ from cell-size trajectories, we approximate $\ln \lambda$ by a linear combination of $M=M_0 \times M_1$ basis functions
\begin{equation}
\ln \lambda(\cs{t}{}, \cs{t}{\ast}) = \sum_{\mu=1}^M w_\mu \vartheta_\mu(\cs{t}{}, \cs{t}{\ast}) 
= \sum_{\mu_0 = 0}^{M_0 - 1}\sum_{\mu_1=0}^{M_1-1}w_{(\mu_0, \mu_1)} {\theta}_{\mu_0}(\cs{t}{}) {\theta}_{\mu_1}^\ast(\cs{t}{\ast}), \label{eqn:superposition}
\end{equation}
where each basis function $\vartheta_\mu$ is decomposed into the product of univariable functions ${\theta}_{\mu_i}^{\ast\cdots\ast}$ and the weights $\mathbf{w}=[w_1, w_2, \dots, w_M]^T$ encode the information about $\lambda$ that we intend to learn from data. To perform Bayesian inference, we aim to minimize the negative log-posterior 
\begin{equation}
-\ln P(\mathbf{w} | \cs{t}{} ) \sim -\ln P(\cs{t}{} | \mathbf{w}) - \ln P(\mathbf{w})
\end{equation}
with respect to $\mathbf{w}$ given the trajectory $\cs{t}{}$. To prevent overfitting, we follow previous work \cite{tipping2001sparse, wipf2004sparse} and impose a sparsity-promoting Gaussian prior over the weights 
\begin{equation}
P(\mathbf{w}) \equiv P(\mathbf{w} | \boldsymbol{\gamma}) 
= \prod_{\mu=1}^M \mathcal{N}(w_\mu | 0, \gamma_\mu)
= \prod_{\mu=1}^M (2\pi \gamma_\mu)^{-1/2} \exp\Big( - \frac{w_\mu^2}{2\gamma_\mu}\Big),
\end{equation}
where $\boldsymbol{\gamma}$ is a vector of $M$ hyperparameters representing the variances of the Gaussian distributions. Thus, the negative log-prior reads
\begin{equation}
-\ln P(\mathbf{w}) = \sum_{\mu=1}^M \frac{w_\mu^2}{2\gamma_\mu} + \sum_{\mu=1}^M \frac{1}{2}\ln(2\pi\gamma_\mu),
\end{equation}
which is similar to an L2 regularization on the weights \cite{krogh1991simple, mackay1992practical}.
The likelihood of the observed trajectory, with a finite number of cell division events occurring at times $\{\tau_j\}$, is given by
\begin{equation}
P(\cs{t}{} | \mathbf{w}) = \exp\Big[-\int \lambda(\cs{t}{}, \cs{t}{\ast}) dt \Big]\prod_{j} \lambda(\cs{t}{}, \cs{t}{\ast})|_{t=\tau_j}. \label{eqn:likelihood_poisson}
\end{equation}
Substituting the expression Eq.~[\ref{eqn:superposition}] for $\lambda$ into Eq.~[\ref{eqn:likelihood_poisson}], we obtain the negative log-likelihood
\begin{equation}
-\ln P(\cs{t}{} | \mathbf{w})= - \sum_\mu w_\mu \sum_{j} \vartheta_\mu( \cs{\tau_j}{}, \cs{\tau_j}{\ast} )   +   \int \exp \Big[ \sum_\mu w_\mu \vartheta_\mu(\cs{t}{}, \cs{t}{\ast}) \Big] dt.
\end{equation}

\subsubsection*{Data processing}
Given a time-series measurement of cell size $\{\cs{t_\alpha}{}\}~(\alpha=1,2,\dots, N)$, or a collection of these measurements, we first determine the time points at which cell division occurs by setting $\{\tau_j\} = \{t_\alpha | \cs{t_{\alpha}}{} - \cs{t_{\alpha+1}}{}  > \dCs \}$ and sorting $\{\tau_j\}$ in ascending order ($\tau_1 < \tau_2 < \dots$). The threshold $\Delta S^*$ is determined by examining the distribution of $\cs{t_{\alpha}}{} - \cs{t_{\alpha+1}}{}$, and we verify the labeled division events by marking them on time-series trajectories (main Figs.~1--3). Subsequently, we obtain $\cs{t}{\ast}$ to be the stepwise function $\cs{t}{\ast} = \sum_j \cs{\tau_j}{} \chi_{t}^{(\tau_j, \tau_{j+1}]}$ where the indicator function $\chi_{t}^{(\tau_j, \tau_{j+1}]}=1$ if $t\in (\tau_j, \tau_{j+1}]$ and $0$ otherwise (see main Fig.~1 for an example).

\subsubsection*{Choice of basis functions}
Given a set of $N$ data points $\{x_\alpha\}~(\alpha=1, 2, \dots, N)$, we construct an orthonormal basis ${\theta}_m(x)$ from the data, such that ${\theta}_{m}(x)$ is a polynomial of degree $m$ and 
\begin{equation}
\langle {\theta}_{m}, {\theta}_{n} \rangle \equiv \int {\theta}_{m} {\theta}_{n} P(x) dx \approx N^{-1}\sum_\alpha {\theta}_{m}(x_\alpha) {\theta}_{n}(x_\alpha) = \delta_{mn}. \label{eqn:orthobasis}
\end{equation}
To ensure numerical stability and accuracy, we employ the modified Gram-Schmidt procedure to build the orthonormal basis \cite{bjorck1994numerics}. 
We verify that this procedure correctly reproduces the Legendre, Hermite, and Chebyshev (first kind) polynomials when applied to synthetic data drawn from uniform, Gaussian, or beta distributions, respectively (Fig.~\ref{SIfig_orthobasis}).
We use this procedure to generate ${\theta}_{\mu_0}$ from $\{ \cs{t_\alpha}{}\}$ and ${\theta}^{\ast}_{\mu_1}$ from $\{\cs{t_\alpha}{\ast}\}$, leading to an $N\times M$ library matrix $\boldsymbol{\Theta}_{\alpha \mu} \equiv \vartheta_\mu(\cs{t_\alpha}{},\cs{t_\alpha}{\ast})$ (see Eq.~[\ref{eqn:superposition}]).

\subsubsection*{Numerical optimization}
Using the library matrix, we can rewrite Eq.~[\ref{eqn:superposition}] as $(\ln \lambda)_{t_\alpha} = \sum_\mu \boldsymbol{\Theta}_{\alpha \mu}\mathbf{w}_\mu$. By defining $\tilde{\boldsymbol{\theta}}_\mu = \sum_{\tau \in \{\tau_j\} } \vartheta_\mu( \cs{\tau}{}, \cs{\tau}{\ast} )$ and using the trapezoidal rule to approximate time integral $\int f(t) dt \approx \sum_\alpha f(t_\alpha) \frac{t_{\alpha+1}-t_{\alpha-1}}{2}\equiv \sum_\alpha f(t_\alpha) \Delta t_\alpha$, we obtain the negative log-posterior
\begin{equation}
-\ln P(\mathbf{w} | \cs{t}{}) \approx -\mathbf{w}_\mu \tilde{\boldsymbol{\theta}}_{\mu} + \Delta t_\alpha \exp(\boldsymbol{\Theta}_{\alpha \mu} \mathbf{w}_\mu) + \frac{\mathbf{w}_\mu^2}{2\boldsymbol{\gamma}_\mu} + c_{\gamma},
\label{eqn:log-posterior}
\end{equation}
where $c_\gamma$ is a constant that doesn't depend on $\mathbf{w}$ and we have used the Einstein summation convention for brevity. For given values of $\boldsymbol{\gamma}$, we use the limited memory Broyden–Fletcher–Goldfarb–Shanno (L-BFGS) algorithm to find the optimal $\hat{\mathbf{w}}$ that minimizes the negative log-posterior Eq.~[\ref{eqn:log-posterior}] \cite{liu1989limited}. The L-BFGS method is a quasi-Newton methods that requires evaluation of the gradient given by
\begin{equation}
-\frac{\partial\ln P(\mathbf{w} | \cs{t}{} )}{\partial \mathbf{w}_\mu} = -\tilde{\boldsymbol{\theta}}_\mu + \Delta t_\alpha \boldsymbol{\Theta}_{\alpha \mu} \exp(\boldsymbol{\Theta}_{\alpha \nu} \mathbf{w}_\nu) + \mathbf{w}_\mu / \boldsymbol{\gamma}_\mu.
\end{equation}
To determine the unknown values of $\boldsymbol{\gamma}$, we adopt a pragmatic procedure based on previous work \cite{mackay1992bayesian}, and choose $\boldsymbol{\gamma}$ to maximize the marginal likelihood $P(\cs{t}{}| \boldsymbol{\gamma}) = \int P(\cs{t}{} | \mathbf{w}) P(\mathbf{w} | \boldsymbol{\gamma})  d\mathbf{w}$. Since such values of $\boldsymbol{\gamma}$ cannot be obtained in close form, we employ the Expectation-Maximization (EM) method for iteratively update the values of $\boldsymbol{\gamma}$. Specifically, given $\boldsymbol{\gamma}^\mathrm{o}$ from the previous iteration, and the corresponding maximum a posteriori (MAP) estimation $\hat{\mathbf{w}}^\mathrm{o} = \mathrm{argmax} P(\mathbf{w}| \cs{t}{}, \boldsymbol{\gamma}^\mathrm{o})$, an EM approach gives the re-estimate $\boldsymbol{\gamma}^\mathrm{n}_\mu = \mathrm{E}[\mathbf{w}_\mu^2]_{P(\mathbf{w}| \cs{t}{}, \boldsymbol{\gamma}^\mathrm{o})}$. The exponential term in Eq.~[\ref{eqn:log-posterior}] makes it challenging to analytically compute the expectation value. To facilitate analysis, we use the Laplace's approximation to expand $\ln P(\mathbf{w}| \cs{t}{}, \boldsymbol{\gamma}^\mathrm{o})$ around the MAP point $\hat{\mathbf{w}}^\mathrm{o}$, which gives $\ln P(\mathbf{w}| \cs{t}{}, \boldsymbol{\gamma}^\mathrm{o}) \approx \ln P(\hat{\mathbf{w}}^\mathrm{o}| \cs{t}{}, \boldsymbol{\gamma}^\mathrm{o}) 
-  \frac{1}{2} (\mathbf{w} - \hat{\mathbf{w}}^\mathrm{o})^T\hat{\mathbf{H}}(\mathbf{w} - \hat{\mathbf{w}}^\mathrm{o})^T$ \cite{mackay1994bayesian, mackay2003information}. 
Here, the Hessian $\hat{\mathbf{H}}$ is defined as $\hat{\mathbf{H}}\equiv -\nabla_\mathbf{w} \nabla_\mathbf{w} \ln P(\mathbf{w} | \cs{t}{}, \boldsymbol{\gamma}^\mathrm{o}) |_{\mathbf{w} =\hat{\mathbf{w}}^\mathrm{o}}$. This is equivalent to approximating the posterior by a Gaussian distribution 
$P(\mathbf{w}| \cs{t}{}, \boldsymbol{\gamma}^\mathrm{o}) \approx \mathcal{N}(\mathbf{w}|\boldsymbol{\mu}=\hat{\mathbf{w}}^\mathrm{o}, \boldsymbol{\Sigma}=\hat{\mathbf{H}}^{-1})$, leading to $\mathrm{E}[\mathbf{w}_\mu^2]_{P(\mathbf{w}| \cs{t}{}, \boldsymbol{\gamma}^\mathrm{o})} \approx (\hat{\mathbf{w}}^\mathrm{o}_\mu)^2 + (\hat{\mathbf{H}}^{-1})_{\mu\mu}$. The Hessian of the log-posterior can be calculated from Eq.~[\ref{eqn:log-posterior}] to be $\hat{\mathbf{H}}_{\mu\nu} = \boldsymbol{\gamma}_\mu^{-1} \delta_{\mu\nu} +  \boldsymbol{\Theta}_{\alpha \mu} \boldsymbol{\Theta}_{\alpha \nu} \exp(\boldsymbol{\Theta}_{\alpha \nu} \mathbf{w}_\nu) \Delta t_\alpha$. 
In practice, we found numerically that $(\hat{\mathbf{H}}^{-1})_{\mu\mu}\sim O(N^{-1})$ where $N$ is the number of data points. Thus, for a sufficiently large amount of data, we use $\boldsymbol{\gamma}^\mathrm{n}_\mu = \mathrm{E}[\mathbf{w}_\mu^2]_{P(\mathbf{w}| \cs{t}{}, \boldsymbol{\gamma}^\mathrm{o})}\approx (\hat{\mathbf{w}}^\mathrm{o}_\mu)^2$ to update the estimate of $\boldsymbol{\gamma}$ during iterations. Finally, following the SINDy framework \cite{brunton2016discovering, kaheman2020sindy, messenger2021weaka, messenger2021weakb}, we sequentially threshold the optimized weights, removing unimportant basis functions from the library $\boldsymbol{\Theta}$, and repeat the above process to generate a series of models with increasing sparsity in $\mathbf{w}$. 

\subsubsection*{Model selection}
To identify the model that best balances the goodness of fit and model complexity, we use a modified Bayesian information criteria (BIC) for model selection. Following the standard derivation of BIC \cite{ghosh2006introduction, bhat2010derivation}, we use $\ln P(\cs{t}{}) \approx \ln P(\cs{t}{} | \hat{\mathbf{w}} ) - 1/2\ln |\hat{\mathbf{H}}|$ to measure the effectiveness of different models, where $|\hat{\mathbf{H}}|$ is the determinant of the Hessian described above. The standard BIC assumes that the measurements are iid and invoke weak law of large numbers to obtain $|\hat{\mathbf{H}}|$. Here, we construct $\hat{\mathbf{H}}$ from the analytical expression above and compute its determinant numerically. To make the metric weakly dependent on trajectory lengths and time units, we employ a normalized BIC score $ \big( \ln P(\cs{t}{} | \hat{\mathbf{w}} ) - 1/2\ln |\hat{\mathbf{H}}|\,\big)/\sum_\alpha \Delta t_\alpha$ such that the score is always $-1$ for the worst zero-term model (main Figs.~2,3).

\subsection{Inference of growth rate $g$ and cut size $h$}
As described above, after identifying the time points $\{\tau_j\}$ at which cell division occurs, we obtain a set of original cell sizes $(s_{t^{-}})_{j}=s_{\tau_j}$ and cut sizes $h_j=s_{\tau_j} -s_{\tau_j+\Delta t}$ (where $\Delta t$ is the time increment between two consecutive measurements) at cell division, and perform linear regression to fit $h(s_{t^-})=h_0 + h_1s_{t^-}$.

Similarly, we use the data in the continuous growth phase $\{s_t| t\in \cup_{j}(\tau_j, \tau_{j+1}]\}$ to infer the growth rate $g$. The instantaneous grow rate at $t_\alpha$ can be approximated by central difference $g_\alpha = \frac{s_{t_{\alpha+1}} - s_{t_{\alpha-1}} }{t_{\alpha+1} - t_{\alpha-1}}$. However, such finite-difference methods tend to amplify noise in the data. To mitigate noise, we fit a fifth-order polynomial model $s_{t_\alpha} = \sum_{n=0}^{5} \kappa_n (t_\alpha)^n$ to each segment of the growth phase, which enables an accurate approximation of the time derivatives $g_\alpha = \sum_{n=1}^5 n \kappa_n (t_\alpha)^{n-1}$. We then apply linear regression on $\{g_\alpha, s_{t_\alpha}\}$ to obtain the relation $g(s_t) = g_0 + g_1 s_t$.

\subsection{Application to cell-division data}
To validate our model inference framework, we first apply it to synthetic data of cell growth and division, where the data-generating model is known. Specifically, we simulate Eq.~(\ref{eqn:SDE}) with exponential growth $g(\cs{t}{}) = g_1 s_t$ and symmetric division $h(s_{t^-}) = s_{t^-} / 2$. The instantaneous rate of cell division is given by 
\begin{equation}
\lambda(\cs{t}{}, \cs{t}{\ast}) = \lambda_{\max}\big[1+\tanh\beta(s_t-\Tilde{s}(s_{t}^{\ast}))\big]/2, \label{eqn:lambda_tanh}
\end{equation}
that transits from $\lambda\approx0$, when the cell size is much smaller than the target division size $\beta(s_t- \Tilde{s}) \ll 0$, to the maximal rate of division $\lambda \approx \lambda_{\max}$ when $\beta(s_t- \Tilde{s}) \gg 0$. As discussed in the main text, the target division size $\tilde{s}$ takes the form of a linear function $\tilde{s}(\cs{t}{\ast}) = c \cs{t}{\ast} + \Delta$, which is capable of describing the ``adder'', ``sizer'', and ``timer'' models. Indeed, our inference framework can accurately recover the ground-truth functions $g(\cs{t}{})$, $h(\cs{t^-}{})$, and $\lambda(\cs{t}{}, \cs{t}{\ast})$ from the synthetic data (Fig.~\ref{SIfig_simulateddata}).

We next apply our inference framework to time-series data of cell sizes in various organisms, including \textit{Escherichia coli} \cite{wang2010robust}, \textit{Schizosaccharomyces pombe} \cite{nakaoka2017aging}, \textit{Dictyostelium discoideum} \cite{tanaka2021dynamic}, \textit{Bacillus subtilis} \cite{nordholt2020biphasic}. Our framework is directly applicable to these datasets and discovers SDE models that quantitatively capture the experimentally measured statistics, including the distributions of division size and generation time, as well as the correlations of cell sizes at two consecutive divisions (main Fig.~2--3 and Figs.~\ref{SIfig_amoeba}--\ref{SIfig_statistics}). The learned models reveal nonlinear memory in cell division dynamics, extending beyond the traditional ``adder'', ``sizer'', and  ``timer'' models.

\subsection{Applications to other classes of systems: website user activities, clinical visits, and earthquakes}
In principle, our Bayesian inference framework should be broadly applicable to learning the discrete-time dynamics from data. Here, we demonstrate its versatility by applying it to three real-world datasets that record a series of discrete events. The first dataset contains the badge-acquisition history of 663 users over a two-year span on \href{https://stackoverflow.com/}{Stack Overflow}, an online question-answering website \cite{du2016recurrent, jia2019neural} (Fig.~\ref{SIfig_stackexchange}). The second dataset contains the clinical visit history of 120 de-identified patients in an Intensive Care Unit (ICU) \cite{du2016recurrent, jia2019neural} (Fig.~\ref{SIfig_icu}). The third dataset records the occurrences of earthquakes with magnitudes of 2.5 or higher at various locations over a thirty-day span \cite{usgs_web} (Fig.~\ref{SIfig_earthquake}). All these data contain a series of time points $\{\tau_j\}$ where certain events occur.

To model a series of discrete events in a framework similar to Eq.~(\ref{eqn:SDE}), we construct a new time series $T_t = t - \sum_j\tau_j \chi_{t}^{(\tau_j, \tau_{j+1}]}$, describing the waiting time $T$ from the previous events. Here, $\chi_{t}$ is the indicator function as before. Accordingly, the dynamics of $T_t$ is governed by the following SDE:
\begin{equation}
\label{eqn:SDE_discrete}
\D T_t =  \D t \, - T_{t^-} \D \mathcal{N}_t(\lambda).
\end{equation}
This is analogous to Eq.~(\ref{eqn:SDE}) that describes cell growth and division with an effective growth rate $g = 1$ and an effective cut size $h(T_{t^-})=T_{t^-}$. 
To incorporate memory in the Poisson intensity $\lambda$, we define another time series $T^*_t=\sum_jT_{\tau_j}\chi_{t}^{(\tau_j, \tau_{j+1}]}=\sum_j (\tau_j-\tau_{j-1})  \chi_{t}^{(\tau_j, \tau_{j+1}]} $, similar to the mother size $\cs{t}{\ast}$ in the case of cell division, to represent the waiting time between two previous events. Subsequently, we can apply our Bayesian inference framework to the transformed time series $T_t$ and $T_t^\ast$ to learn a Poisson intensity $\lambda(T_t, T_t^\ast)$ that encodes the discrete dynamics. As shown in Figs.~\ref{SIfig_stackexchange}--\ref{SIfig_earthquake}, our framework identifies sparse models that capture the ensemble statistics in all three input datasets. These results demonstrate the broad applicability of our model discovery pipeline to real-world datasets, particularly in fields like healthcare and geoscience.

\subsection{Robustness of the inference framework\label{sec:discussion}}
In this subsection, we demonstrate the robustness and adaptability of our inference framework.
We start by examining the performance of our framework using alternative basis functions in Eq.~(\ref{eqn:superposition}) for approximating $\lambda(\cs{t}{}, \cs{t}{\ast})$. Specifically, we test two additional basis: a local Gaussian kernel basis  $\theta_\mu(s; s_\mu, \sigma) = \exp[\frac{-(s-s_\mu)^2}{2\sigma^2}]$, and a log-sigmoid basis $\theta_\mu(s; s_\mu) = -\ln[1+\exp(-(s-s_\mu))] + 1$ inspired by the asympototic behavior of Eq.~(\ref{eqn:lambda_tanh}). Both choices of basis functions can be readily implemented in our framework, and yield accurate representation o the ground-truth $\lambda$ when applied to the simulation dataset (Fig.~\ref{SIfig_varbasis}).

Moreover, our framework is robust against different regularizers for promoting sparsity. In the Bayesian framework, our goal is to minimize the negative log-posterior Eq.~(\ref{eqn:log-posterior}) where the log-Gaussian-prior $\sum_\mu \frac{w_\mu^2}{2\gamma_\mu}$ serves as the regularization term. Similarly, we can introduce different regularization terms to Eq.~(\ref{eqn:log-posterior}), such as $\lambda_1 \sum_\mu|w_\mu|$ for lasso regularization, $\lambda_2 \sum_\mu w_\mu^2$ for ridge regularization, and $\lambda_1 \sum_\mu|w_\mu| + \lambda_2 \sum_\mu w_\mu^2$ in the elastic net (L1 + L2) methods. As shown in Fig.~\ref{SIfig_varreg}, the inference results are largely insensitive to the choice of regularizers and all of them are able to reproduce the ground-truth model.

Furthermore, our framework can interface with deep learning setups, which may enhance the inference capabilities by employing the well-developed machine learning (ML) libraries (such as PyTorch and TensorFlow). To integrate ML into our framework, we use a multi-layer perceptron neural network to approximate the Poisson intensity $\lambda(s_t, s_t^\ast)$ and a mini-batch Adam optimizer to minimize the negative log-posterior. This setup yields similar inference results as our Bayesian framework (Fig.~\ref{SIfig_varbasis}D). A Google Colab notebook is included in our \href{https://github.com/f-chenyi/cell_growth}{Github repository} to demonstrate this example.

\section{Analysis of SDE models of cell growth and division\label{sec:S2}}
\subsection{Analysis of the no-memory model\label{sec:memoryless_analysis}}

We start by looking at the no-memory model with an instantaneous cell division rate $\lambda = \lambda(\cs{t}{})$.
To analyze the probability distribution of cell size at division $\cs{}{\ast}$ and that of generation time $\tau$,
we consider a simple version of Eq.~(\ref{eqn:SDE}) where $g(\cs{t}{}) = \cs{t}{}$, describing an exponential growth (time normalized by growth rate), $h(\cs{t^-}{}) = \cs{t^-}{}/2$, describing a symmetric cell division, and $\lambda(\cs{t}{}) = \cs{t}{}^2$. A typical simulated trajectory is shown in Fig.~\ref{SIfig_nomemory}A. Note that we use the same notation $\cs{}{\ast}$ for ``mother size" in the section above because it is the cell size at previous division.

The probability density $p(\cs{}{\ast})$ is given by 
$
p(\cs{}{\ast}) = \int_0^{\infty} P(\cs{}{\ast}| \cs{0}{}) q(\cs{0}{}) \D s_0,
$
where $q(\cs{0}{})$ is the marginal probability of the cell size $\cs{0}{}$ at birth, $P(\cs{}{\ast}| \cs{0}{})$ is the conditional probability of dividing at $\cs{}{\ast}$ given the born size $\cs{0}{}$. To solve for $p(\cs{}{\ast})$, we note that $p(\cs{}{\ast})$ and $q(\cs{0}{})$ are related by $q(\cs{0}{}) = 2p(2\cs{0}{})$ due to symmetric division. Introducing this relation into the expression for $p(\cs{}{\ast})$, we obtain an integral equation for $p$
\begin{equation}
p(\cs{}{\ast}) = 2\int_0^{\infty} P(\cs{}{\ast}| \cs{0}{}) p(2 \cs{0}{}) \D \cs{0}{}. \label{eqn:integral_symdiv}
\end{equation}
To compute $P(\cs{}{\ast}| \cs{0}{})$, we first calculate the probability $\mathcal{P}(\cs{}{\ast}| \cs{0}{})$ of dividing at a cell size larger than $\cs{}{\ast}$ given a born size of $\cs{0}{}$, which is given by
\begin{equation}
\mathcal{P}(\cs{}{\ast}| \cs{0}{}) = \begin{cases}
1 & ~\mathrm{when}~\cs{}{\ast} < \cs{0}{}, ~\mathrm{and}\\
\exp\Big( -\int_0^{T(\cs{}{\ast},\cs{0}{})} \lambda(\cs{t}{}) \D t \Big)  & ~\mathrm{when}~\cs{}{\ast} \geqslant \cs{0}{}.
\end{cases}
\label{eqn:jump_rate_general}
\end{equation}
Here, $T(\cs{}{\ast},\cs{0}{})$ is the generation time of a cell that begins with a size $\cs{0}{}$ at birth and divide at $\cs{}{\ast}$.
In the case of $g(\cs{t}{}) = \cs{t}{}$ and $\lambda(\cs{t}{}) = \cs{t}{}^2$, we obtain $T(\cs{}{\ast},\cs{0}{}) = \ln(\cs{}{\ast}/\cs{0}{})$, $\lambda(\cs{t}{}) = \cs{0}{}^2 e^{2t}$, and hence $\mathcal{P}(\cs{}{\ast} | \cs{0}{}) = \exp(-\frac{(\cs{}{\ast})^2-\cs{0}{}^2}{2})$ when $\cs{}{\ast} > \cs{0}{}$. Since $P(\cs{}{\ast} | \cs{0}{})$ is given by $P(\cs{}{\ast} | \cs{0}{}) =-\frac{\D \mathcal{P}(\cs{}{\ast}|\cs{0}{})}{\D \cs{}{\ast}} $, we obtain 
\begin{equation}
P(\cs{}{\ast} | \cs{0}{}) = \begin{cases}
0 & ~\mathrm{when}~\cs{}{\ast} < \cs{0}{}, ~\mathrm{and}\\
\cs{}{\ast} \exp(-\frac{(\cs{}{\ast})^2-\cs{0}{}^2}{2}) & ~\mathrm{when}~\cs{}{\ast} \geqslant \cs{0}{}. \label{eqn:condprob_quadratic}
\end{cases}
\end{equation}
Introducing Eq.~(\ref{eqn:condprob_quadratic}) into Eq.~(\ref{eqn:integral_symdiv}), we obtain an integral equation
\begin{equation}
p(\cs{}{\ast}) = 2 \cs{}{\ast} \exp(-(\cs{}{\ast})^2/2) \int_0^{\cs{}{\ast}}\exp(\cs{0}{}^2/2)p(2\cs{0}{}) \D \cs{0}{}. \label{eqn:integral_symdiv_quad}
\end{equation}
By inserting an ansatz $p(x) = C(x^2) x \exp(-x^2/2)$ into Eq.~(\ref{eqn:integral_symdiv_quad}), we obtain 
$
C(x^2) = 2\int_0^x \exp(y^2/2)C(4y^2)2y\exp(-2y^2)dy = 2\int_0^{x^2}\exp(-3z/2)C(4z) dz.
$
Differentiating this equation with respect to $x^2$ leads to a {\it functional differential equation} for $C(z)$
\begin{gather}
C'(z) = 2\exp(-3z/2) C(4z), ~\label{eqn:fde_quadratic}
\end{gather}
which can be solved iteratively by $C(z) = \sum_{n=0}^{\infty} C_n(z)$ where $C_0\equiv 1$ and $C_n^\prime(z) = 2\exp(-3z/2) C_{n-1}(4z)$. The recurrence differential equation yields $C_n(z) = \frac{4^n}{(4,4)_n}\exp(-\frac{4^n-1}{2}z)$ where $(q,q)_n = \Pi_{k=1}^n(1-q^k)$ with $(q,q)_0=1$ is the $q$-Pochhammer symbol, and the first few $C_n(z)$ are given by 
\begin{equation}
C_0(z) = 1,~C_1(z) = -\frac{4}{3}\exp(-\frac{3}{2}z),~C_2(z) = \frac{16}{45}\exp(-\frac{15}{2}z), ~C_3(z) = \frac{64}{2835}\exp(-\frac{63}{2}z), \dots
\end{equation}
Finally, plugging $C_n$ into the ansatz, we obtain
\begin{equation}
p(x) =  N^{-1} \sum_{n=0}^{\infty} \frac{4^n}{(4,4)_n} x\exp(-2^{2n-1}x^2) \equiv N^{-1} \sum_{n=0}^{\infty} p_n(x),
\end{equation}
where $N$ is the normalization constant given by
$$
N = \sum_{n=0}^\infty \int_0^\infty p_n(x) \D x = \sum_{n=0}^\infty \frac{1}{(4,4)_n}.
$$ 
As shown in Fig.~\ref{SIfig_nomemory}B, our analytical results are in good agreement with the simulation distribution.

Next, to compute the distribution $\tilde{p}(\tau)$ of generation time $\tau$, we can replace the conditional probability $P(\cs{}{\ast}|\cs{0}{})$ in Eq.~(\ref{eqn:integral_symdiv}) with the conditional probability $\td{P}(\tau | \cs{0}{})$ of having a generation time $\tau$ given the born size $\cs{0}{}$, and then obtain $\td{p}(\tau)$ from 
$
\td{p}(\tau) = 2\int_0^{\infty} \tilde{P}(\tau | \cs{0}{}) p(2 \cs{0}{}) \D \cs{0}{}.
$
To obtain $\td{P}(\tau | \cs{0}{})$, we repeat the same process as above. Specifically, we derive the probability $\td{\mathcal{P}}(\tau | \cs{0}{})$ of having a generation time longer than $\tau$ to be $\td{\mathcal{P}}(\tau | \cs{0}{}) = \exp[-\cs{0}{}^2(e^{2\tau}-1)/2]$, and thus $\td{P}(\tau | \cs{0}{}) = -\frac{\D \td{\mathcal{P}}(\tau | \cs{0}{})}{\D \tau}= \cs{0}{}^2e^{2\tau} \exp[-\cs{0}{}^2(e^{2\tau}-1)/2]$. Using this expression for $\td{P}(\tau | \cs{0}{})$ and the expression of $p(x)$, we obtain $\tilde{p}(\tau_d) = f(e^{2\tau})$, where
\begin{align}
f(z) &= 2N^{-1}\int_0^\infty x^2 z \exp(-x^2z/2) \exp(x^2/2) 2x \sum_n \frac{4^n}{(4,4)_n}\exp(-2^{2n+1}x^2) dx \notag \\
& = 2z N^{-1}\sum_{n=0}^{\infty} \frac{4^n}{(4,4)_n} \int_0^\infty y \exp(-yz/2+y/2-2^{2n+1} y) dy \notag \\
& = 2z N^{-1}\sum_{n=0}^{\infty} \frac{4^{n+1}}{(4,4)_n(4^{n+1}-1 + z)^2}.
\end{align}
Indeed, the analytical results agree almost perfectly with the simulation distribution (Fig.~\ref{SIfig_nomemory}C).

Another theoretically interesting case is when $P(\cs{}{\ast} | \cs{0}{})$ is (almost) independent of $\cs{0}{}$. Consider an intensity
$$
\lambda(\cs{t}{}) = \begin{cases}
\alpha \cs{t}{} (\cs{t}{} - \cs{\mathrm{c}}{}) & \cs{t}{} > \cs{\mathrm{c}}{}\\
0 & \cs{t}{} < \cs{\mathrm{c}}{},
\end{cases}
$$ 
where $\cs{\mathrm{c}}{}$ denotes a critical cell size below which the division rate $\lambda$ is zero regardless of born size $\cs{0}{}$.
Introducing this expression for $\lambda$ into Eq.~(\ref{eqn:jump_rate_general}) yields 
\begin{gather}
\mathcal{P}(\cs{}{\ast} | \cs{0}{}) = \begin{cases}
    1 & ~\mathrm{when~} \cs{}{\ast} < \cs{\mathrm{c}}{} ~\mathrm{or}~ \cs{}{\ast} <\cs{0}{},\\
    \exp\big[ -\frac{\alpha}{2}(\cs{}{\ast}-\cs{\mathrm{c}}{})^2\big] & ~\mathrm{when~} \cs{}{\ast} > \cs{\mathrm{c}}{} ~\mathrm{and}~ \cs{0}{}<\cs{\mathrm{c}}{},~\mathrm{and}\\
    \exp\big[ \frac{\alpha}{2}(\cs{}{\ast}-\cs{0}{})(\cs{}{\ast}+\cs{0}{}-2\cs{\mathrm{c}}{})\big]& ~\mathrm{when~} \cs{}{\ast} > \cs{\mathrm{c}}{} ~\mathrm{and}~ \cs{0}{}>\cs{\mathrm{c}}{}.
\end{cases}
\end{gather}
When $\alpha$ is large enough so that (almost) all the cells will divide before they reach a length of $2\cs{\mathrm{c}}{}$, then $\cs{0}{}$ must be smaller than $\cs{\mathrm{c}}{}$, and thus $P(\cs{}{\ast} | \cs{0}{}) \approx \alpha(\cs{}{\ast} - \cs{\mathrm{c}}{}) \exp\big[ -\frac{\alpha}{2}(\cs{}{\ast} -\cs{\mathrm{c}}{})^2\big]~(\forall~\cs{}{\ast} > \cs{0}{})$ – independent of $\cs{0}{}$! Consequently, $p(\cs{}{\ast})$ should be same as $P(\cs{}{\ast} |\cs{0}{})$. One can verify using Eq.~(\ref{eqn:integral_symdiv}) that indeed $p(\cs{}{\ast}) = \alpha(\cs{}{\ast} - \cs{\mathrm{c}}{}) \exp\big[ -\frac{\alpha}{2}(\cs{}{\ast} -\cs{\mathrm{c}}{})^2\big] \int_0^\infty p(2\cs{0}{}) \D (2\cs{0}{}) = \alpha(\cs{}{\ast} - \cs{\mathrm{c}}{}) \exp\big[ -\frac{\alpha}{2}(\cs{}{\ast} -\cs{\mathrm{c}}{})^2\big]$, where the last step uses the basic property of a probability distribution $\int_0^\infty p(x) dx = 1$. After obtaining $p(\cs{}{\ast})$, one can repeat the same process as above to derive $P(\tau | \cs{0}{})$ and compute $\tilde{p}(\tau)$ from $P(\tau | \cs{0}{})$ and $p(\cs{0}{})$. 
These results are also verified by our simulations (Fig.~\ref{SIfig_nomemory}, \textit{bottom}).

\subsection{Analysis of the one-generation memory model\label{sec:memory_analysis}}

To capture potential memory in cell division, we also considered an instantaneous division rate $\lambda(\cs{t}{},\cs{t}{\ast})$ that depends not only on the current cell size $\cs{t}{}$ but also on the cell size $\cs{t}{\ast}$ at the previous division. 
In this case, to analyze the probability density $p(\cs{}{\ast})$ of division size $\cs{}{\ast}$, we obtain an integral equation similar to Eq.~(\ref{eqn:integral_symdiv})
\begin{equation}
p(\cs{}{\ast}) = \int P(\cs{}{\ast} | \cs{}{{\ast\ast}}) p(\cs{}{{\ast\ast}}) \D \cs{}{{\ast\ast}}, \label{eqn:integral_memory}
\end{equation}
where $P(\cs{}{\ast} |\cs{}{{\ast\ast}})$ denotes the conditional probability of dividing at $\cs{}{\ast}$ given the cell size $\cs{}{{\ast\ast}}$ at the last cell division, and we have used the fact that both $\cs{}{\ast}$ and $\cs{}{{\ast\ast}}$ follow the same distribution. To facilitate analysis, we again consider an exponential growth $g(\cs{t}{}) = \cs{t}{}$ with symmetric division $h(\cs{t^-}{}) = \cs{t^-}{} / 2$ and an inhomogeneous intensity 
\begin{equation}
\label{eqn:lambda_analysis_memory}
\lambda(\cs{t}{}, \cs{t}{\ast}) = 
\begin{cases}
0 & \cs{t}{}  < \bar{s}_t = \phi \cs{\mathrm{c}}{} + (1-\phi) \cs{t}{\ast}\\
\alpha \cs{t}{} (\cs{t}{} - \bar{s}_t) & \cs{t}{} \geqslant \bar{s}_t
\end{cases}.
\end{equation}
When $\phi = 1$, $\lambda(\cs{t}{}, \cs{t}{\ast})$ becomes independent of $\cs{t}{\ast}$, similar to a sizer model without memory; when $\phi = 0$, the model does not have an intrinsic target size ($\cs{\mathrm{c}}{}$) for division, similar to a timer model. Given the growth and division rules, we can repeat the same procedures as in Sec.~2\ref{sec:memoryless_analysis}, which yields 
$$
P( \cs{}{\ast} | \cs{}{{\ast\ast}}) = \begin{cases}
0 & \cs{}{\ast} < \bar{s}= \phi\cs{\mathrm{c}}{} + (1-\phi) \cs{}{{\ast\ast}} \\
\alpha (\cs{}{\ast} - \bar{s}) \exp\big[ -\alpha (\cs{}{\ast} - \bar{s})^2 / 2 \big] & \cs{}{\ast} \geqslant  \bar{s}
\end{cases}
$$  
Thus, the probability density $p(\cs{}{\ast})$ is determined by the following integral equation
\begin{equation}
p(\cs{}{\ast}) = \int_0^{\frac{\cs{}{\ast} - \phi \cs{\mathrm{c}}{}}{1-\phi}}  \alpha (\cs{}{\ast} - \bar{s}) e^{-\alpha (\cs{}{\ast} - \bar{s})^2 / 2} p(\cs{}{{\ast\ast}}) \D \cs{}{{\ast\ast}} \label{eqn:integral_memory2}
\end{equation}
where again $\bar{s}$ is given by the weighted average $\bar{s} = \phi \cs{\mathrm{c}}{} + (1-\phi) \cs{}{{\ast\ast}}$. We solved Eq.~(\ref{eqn:integral_memory2}) numerically (via iterations), and verified that the numerical solution $p(\cs{}{\ast})$ agrees well with the distribution generated by simulating the underlying SDE model (Fig.~\ref{SIfig_analysis_memory}B). Furthermore, we can compute the joint distribution $P(l_{\D,-1}, l_\D)$ as $$
P(\cs{}{\ast},\cs{}{{\ast\ast}}) = 
\begin{cases}
\alpha (\cs{}{\ast} - \bar{s}) e^{-\alpha (\cs{}{\ast} - \bar{s})^2 / 2} p(\cs{}{{\ast\ast}})  & \cs{}{{\ast\ast}} < \frac{ \cs{}{\ast} - \phi \cs{\mathrm{c}}{}}{1-\phi} \\
0 & \cs{}{{\ast\ast}}\geqslant \frac{ \cs{}{\ast} - \phi \cs{\mathrm{c}}{}}{1-\phi}
\end{cases}
$$ which is indeed in good agreement with the one obtained from direct simulation of the model (Fig.~\ref{SIfig_analysis_memory}C,~D).

\subsection{Analysis of cell-division memory using singular-value decomposition\label{sec:svd_analysis}}
To visualize whether cell division possesses memory, we have generated the joint probability distribution $P(\cs{}{\ast}, \cs{}{{\ast\ast}})$ of the cell size at two consecutive divisions (see main Figs.~2, 3).
To quantify cellular memory in cell division, we examine the singular-value decomposition (SVD) of $P(\cs{}{\ast},\cs{}{{\ast\ast}}) = \sum_i \Lambda_i u_i(\cs{}{\ast}) v_i(\cs{}{{\ast\ast}})$, where $\Lambda_i$ denotes the singular values (SVs), and $u_i$ and $v_i$ are orthonormal functions that satisfy 
$$
\left< u_i, u_j \right> \equiv \int_{-\infty}^\infty \D x\, u_i(x)u_j(x) = \delta_{ij}
\qquad\quad 
\text{and}
\qquad\quad 
\left< v_i, v_j \right> \equiv \int_{-\infty}^\infty \D y\, v_i(y)v_j(y) = \delta_{ij}
$$ 
The singular value spectrum $\{\Lambda_i\}$ indicates the number of independent ``modes'' (marginal products) needed to reconstruct $P$ and their relative importance. If there is no memory in cell division, meaning that all the division events are independent, then $P(\cs{}{\ast}, \cs{}{{\ast\ast}}) = p(\cs{}{\ast})p(\cs{}{{\ast\ast}})$ is the product of the marginals, and there is only one non-zero SV, corresponding to the decomposition with $u_1(x) = v_1(x) = p(x)/\sqrt{\left< p, p \right>}$. Our analysis of the experimental data (main Figs.~2, 3) shows that the measured joint probability $P(\cs{}{\ast}, \cs{}{{\ast\ast}})$ can have multiple nonzero SVs, suggesting strong memory of cell division.

To understand quantitatively how the nonzero SVs are related to cell-division memory, we study analytically the singular value spectrum of a multivariate normal distribution 
\begin{equation}
P_n(x,y) = (\pi \sigma_+ \sigma_-)^{-1} \exp\left[ -\frac{(x-y)^2}{2 \sigma_-^2} - \frac{(x+y-2\mu)^2}{2 \sigma_+^2} \right],
\label{eqn:multivariate_gaussian}
\end{equation}
where $\sigma_+$ and $\sigma_-$ denote the spread of the probability distribution along and perpendicular to the $y=x$ line. We consider a series of $P_n$ with the same marginal distribution 
$\int_{-\infty}^\infty dy\, P_n(x,y) $, which dictates that $\sigma_+^2 + \sigma_-^2 = \mathrm{const.}$, but with varying ratio $r = \sigma_+ / \sigma_-$. Since $P_n(x,y) = P_n(y, x)$ the singular basis functions $u_i(x)$ and $v_i(x)$ must be the same. One can use the Gram-Schmidt process to construct the orthogonal basis functions, the first few of which are listed below:
\begin{align}
u_1(x) &=\pi^{-1/4}\sigma^{-1/2}\exp(-\frac{x^2}{2\sigma^2}), \quad\notag\\
u_2(x) 
&= 2^{1/2}\pi^{-1/4}\sigma^{-3/2} x \exp(-\frac{x^2}{2\sigma^2})\notag \\
u_3(x) &= 2^{1/2}\pi^{-1/4}\sigma^{-5/2}x^2\exp(-\frac{x^2}{2\sigma^2})- (1/2)^{1/2} u_1(x) \quad\notag\\
u_4(x) &= (4/3)^{1/2}\pi^{-1/4}\sigma^{-7/2}x^3\exp(-\frac{x^2}{2\sigma^2})- (3/2)^{1/2} u_2(x)\notag \\
&\vdots \label{eqn:gs_u}
\end{align}
The parameter $\sigma$ is determined by maximizing the largest SV $\Lambda_1=\left<P_n(x,y), u_1(x-\mu) u_1(y-\mu)\right>$, which yields $\sigma = \sqrt{\sigma_+\sigma_-/2}$. Introducing this expression for $\sigma$ into Eq.~(\ref{eqn:gs_u}), we obtain that $\Lambda_i = \left<P_n(x,y), u_i(x-\mu) u_i(y-\mu)\right> = \Lambda_1 (\frac{r-1}{r+1})^{i-1}~(i=1,2,\dots)$. This analytical expression for $\Lambda_i$ is verified numerically (Fig.~\ref{SIfig_analysis_svd}) at varying values of $r$. Thus, the second largest SV, relative to the largest SV, $\Lambda_2/\Lambda_1 = \frac{r-1}{r+1}$ is closely related to the correlation of cell sizes at two consecutive divisions, which is given by $\frac{r^2-1}{r^2+1}$ for the normal distribution Eq.~(\ref{eqn:multivariate_gaussian}).

\clearpage

\begin{figure*}
\centering
\includegraphics[width=.6\textwidth]{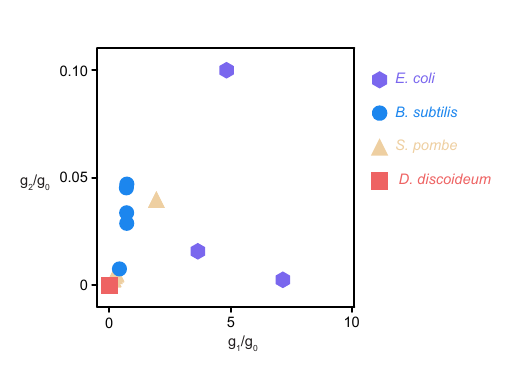}
\caption{Scatter plot of the relative quadratic coefficient $g_2/g_0$ vs. the relative linear coefficient $g_1/g_0$ for the growth rate function $g(s_t) = g_0 + g_1 s_t + g_2 s_t^2$. 
The quadratic terms are negligible for the experimental data in \textit{E.~coli} \cite{wang2010robust}, \textit{B. subtilis} \cite{nordholt2020biphasic}, \textit{S.~pombe} \cite{nakaoka2017aging}, \textit{D. discoideum} \cite{tanaka2021dynamic}.}
\label{SIfig_g_coefs}
\end{figure*}

\clearpage

\begin{figure*}
\centering
\includegraphics[width=.9\textwidth]{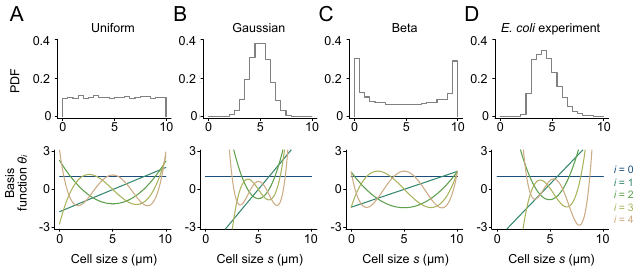}
\caption{Polynomial basis functions that are orthogonal with respect to data. 
Probability distributions of the synthetic or experimental cell-size data (\textit{Top}) and the corresponding orthogonal polynomial basis $\theta_i$ (\textit{Bottom}; see Eq.~\ref{eqn:orthobasis}) are shown the designated datasets. Cell size $s$ are drawn from (\textit{A}) a uniform distribution $\mathcal{U}_{[0,10]}$, (\textit{B}) a Gaussian distribution $\mathcal{N}(\mu=5,\sigma=1)$, (\textit{C}) a beta distribution $\mathcal{B}(\alpha=0.5, \beta=0.5)$, and (\textit{D}) the measured distribution from the \textit{E.~coli} experiments in \cite{wang2010robust}. For \textit{A}--\textit{C}, 10000 data points are generated from each of the underlying distributions. The modified Gram-Schmidt procedure recovers (up to a constant normalization factor) the Legendre polynomials in \textit{A}, the Hermite polynomials in \textit{B}, and the Chebyshev polynomials of the first kind in \textit{C}.  
}
\label{SIfig_orthobasis}
\end{figure*}

\clearpage

\begin{figure*}
\centering
\includegraphics[width=.9\textwidth]{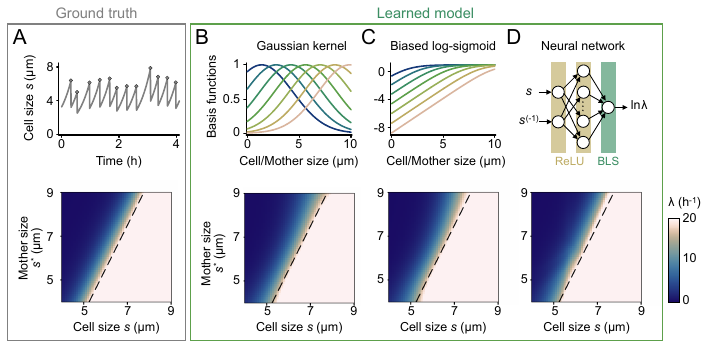}
\caption{The Bayesian inference framework is robust against different choices of function approximation. 
(\textit{A}) A typical trajectory (\textit{Top}) and the ground-truth cell division rate $\lambda$ (\textit{Bottom}) of the synthetic ``adder" model as in main Fig.~1.
(\textit{B} -- \textit{D}) Illustration of different methods of function approximation (\textit{Top}) and the corresponding cell division rate $\lambda$ (\textit{Bottom}) inferred from the input data in \textit{A} (1500 division events in total). We tested (B) Gaussian kernels $\theta_i(s; s_i, \sigma) = \exp[\frac{-(s-s_i)^2}{2\sigma^2}]$ or (C) Biased log-sigmoid functions $\theta_i(s; s_i) = -\ln[1+\exp(-(s-s_i))] + 1$ as alternative basis functions in our Bayesian inference framework. In addition, we also tested using (\textit{D}) a three-layer neural network to approximate $\ln \lambda$ which yielded similar inference results as the basis-function methods. (see Sec.~1\ref{sec:discussion} for details).
}
\label{SIfig_varbasis}
\end{figure*}

\clearpage

\begin{figure}
\centering
\includegraphics[width=\textwidth]{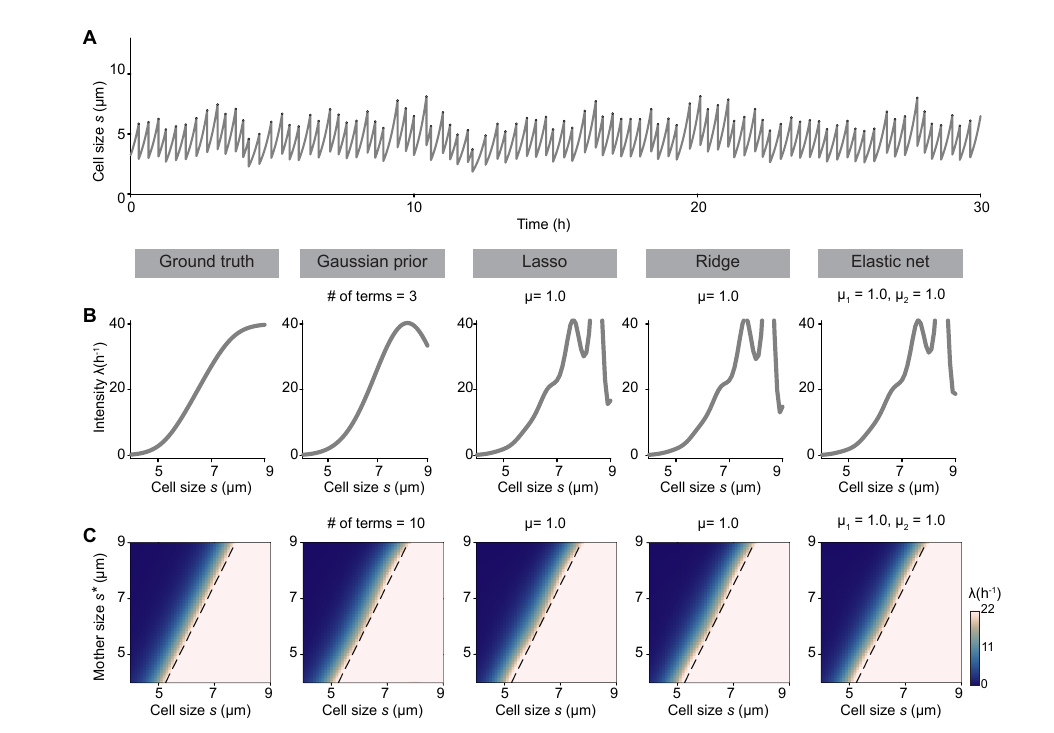}
\caption{
The Bayesian inference framework is robust against different choices of regularization. 
(\textit{A}) A simulated trajectory of cell sizes from the ``adder'' model as in Fig.~1 and \ref{SIfig_simulateddata}. 
(\textit{B}) The ground-truth division rate $\lambda$ averaged over mother size $\cs{t}{\ast}$ and the division rate $\lambda$ of a no-memory model learned with the sparsity-promoting Gaussian prior and the Lasso, Ridge and elastic net regression models.
(\textit{C}) The ground-truth division rate $\lambda$ and the division rate $\lambda$ of a model with one-generational memory learned with the sparsity-promoting Gaussian prior and Lasso, Ridge and elastic net regression model.
}
\label{SIfig_varreg}
\end{figure}

\clearpage

\begin{figure*}
\centering
\includegraphics[width=\textwidth]{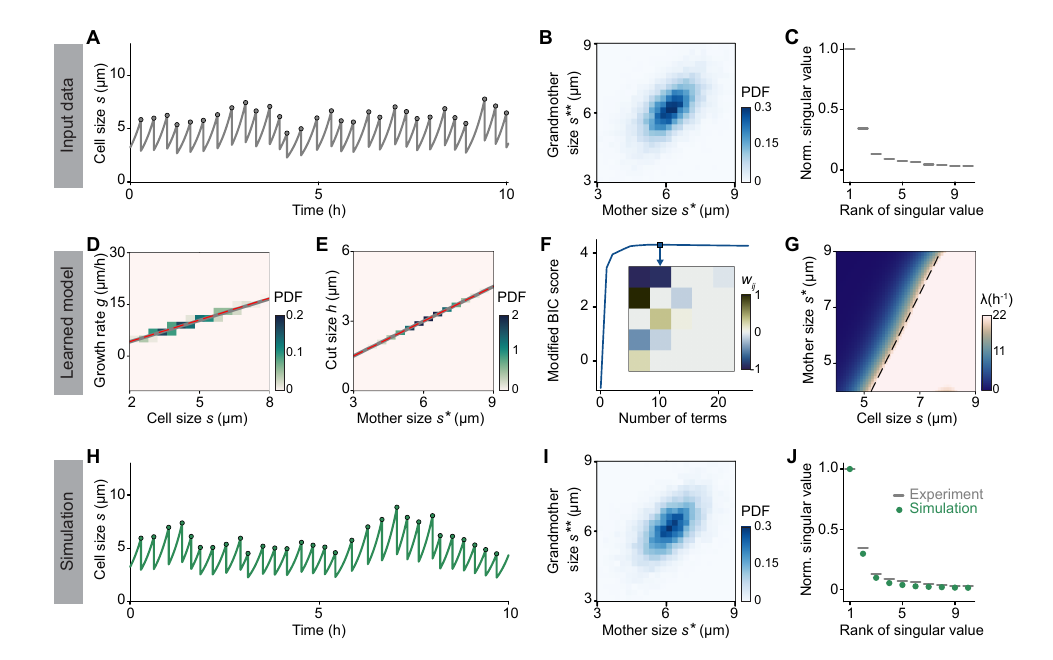}
\caption{The growth and division model learned from simulation data successfully reproduces the ground-truth. 
(\textit{A}) A typical trajectory of cell size from simulations of Eq.~(\ref{eqn:SDE}). The model describes a growth rate $g(s_t)=g_1s_t= (\ln2/\tau)s_t$ with a doubling time $\tau=20~\mathrm{min}$, and a symmetric division $h(s_{t^{-}})=s_{t^{-}}/2$. The division rate is given by Eq.~(\ref{eqn:lambda_tanh}) with parameters $\lambda_{\max}=40~\mathrm{h^{-1}}, \beta = 1.25~\um^{-1}$, and $\tilde{s}(\cs{t}{\ast})=(\cs{t}{\ast}+6.5)/2$.
(\textit{B}) The joint probability distribution of mother size $\cs{}{\ast}$ (cell size at last division) and grandmother size $s^{\ast\ast}$ (cell size at the second last division) shows a strong positive correlation between the sizes of two consecutive divisions. This correlation arises because the target division size $\tilde{s}$ depends on the mother size $\cs{}{\ast}$. 
(\textit{C}) The 10 largest singular values of the joint probability distribution in \textit{B}. Multiple singular values are significantly greater than 0, indicating the correlation between $\cs{}{\ast}$ and $s^{\ast\ast}$ in \textit{B}. 
(\textit{D}) Heatmap of cell size $s$ versus growth rate $g$ from input data. The dashed red line, representing the learned growth rate $g(s)$, overlaps with the solid gray line, which represents the ground truth. 
(\textit{E}) Heatmap of mother size $s^{\ast}$ versus cut size $h$ from input data. The dashed red line, representing the learned cut size $h(\cs{}{\ast})$, overlaps with the solid gray line, which represents the ground truth.
(\textit{F}) Modified Bayesian information criterion (BIC) scores of models with varying number of terms. The square marker indicates the selected model with the highest BIC score, whose coefficients $\mathbf{w}$ are shown in the inset.
(\textit{G}) The learned division rate $\lambda$,  corresponding to the selected model in \textit{F}, faithfully reproduces the ground truth. The black dashed line represents $s = \tilde{s}(\cs{}{\ast})$, indicating the boundary between high and low division rates.
(\textit{H}) -- (\textit{J}) Simulation results of the selected learned model described by \textit{D} -- \textit{G}. Plots correspond to \textit{A} -- \textit{C}, respectively. 
}
\label{SIfig_simulateddata}
\end{figure*}

\clearpage

\begin{figure*}
\centering
\includegraphics[width=.9\textwidth]{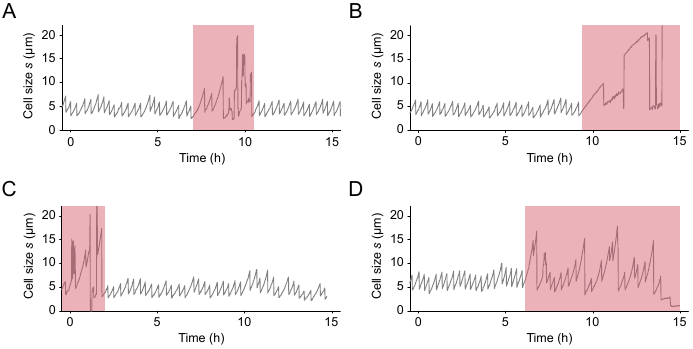}
\caption{The preprocessing steps discard segments of cell-size trajectories that exhibit anomalous growth and division (highlighted in red).  Examples are shown for the experimental data in \textit{E.~coli} \cite{wang2010robust}.}
\label{SIfig_preprocessing}
\end{figure*}

\clearpage

\begin{figure*}
\centering
\includegraphics[width=\textwidth]{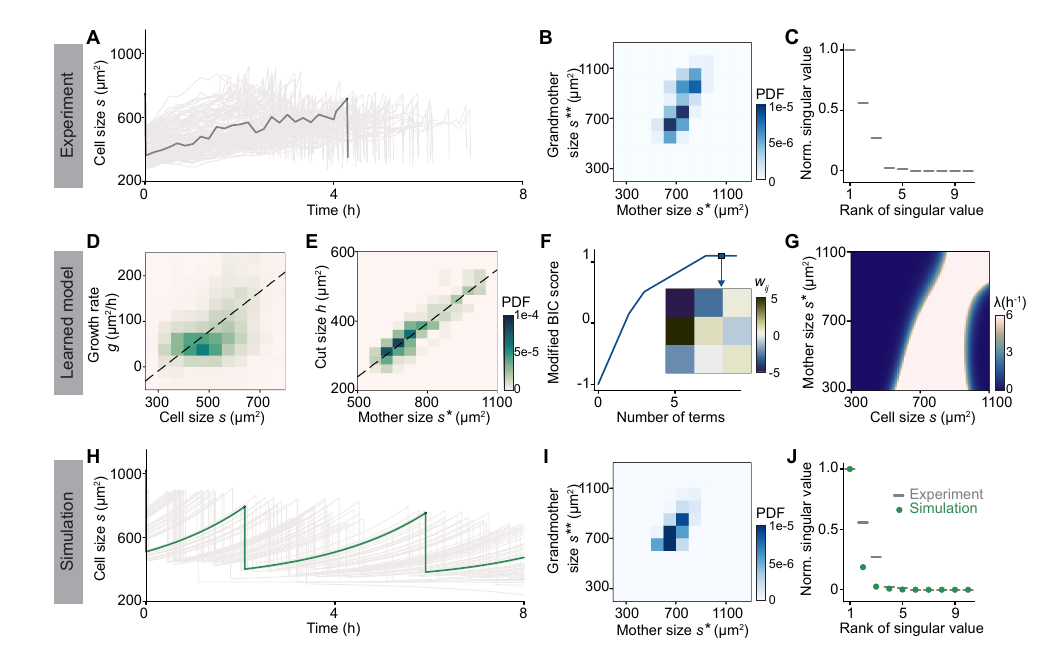}
\caption{The growth and division model learned from experimental data in \textit{D.~discoideum} \cite{tanaka2021dynamic} reveals a strong memory of mother size in cell division. 
(\textit{A}) 147 experimental trajectories of cell sizes are shown in light gray. A typical trajectory is highlighted in dark gray. The doubling time is approximately 4~h. 
(\textit{B}) The joint probability distribution of mother size $\cs{}{\ast}$ (cell size at last division) and grandmother size $s^{\ast\ast}$ (cell size at the second last division) shows a strong positive correlation between the sizes of two consecutive divisions.
(\textit{C}) The 10 largest singular values of the joint probability distribution in \textit{B}. Multiple singular values are significantly greater than 0, indicating the correlation between $\cs{}{\ast}$ and $s^{\ast\ast}$ in \textit{B}. 
(\textit{D}) Heatmap of cell size $s$ versus growth rate $g$ from input data. The dashed line shows the learned growth rate $g(s)$. 
(\textit{E}) Heatmap of mother size $s^{\ast}$ versus cut size $h$ from input data. The dashed line shows the learned cut size $h(\cs{}{\ast})$.
(\textit{F}) Modified Bayesian information criterion (BIC) scores of models with varying number of terms. The square marker indicates the selected model with the highest BIC score, whose coefficients $\mathbf{w}$ are shown in the inset.
(\textit{G}) The learned division rate $\lambda$ corresponding to the selected model in \textit{F}.
(\textit{H}) -- (\textit{J}) Simulation results of the selected learned model described by \textit{D} -- \textit{G}. Plots correspond to \textit{A} -- \textit{C}, respectively. 
}
\label{SIfig_amoeba}
\end{figure*}

\clearpage

\begin{figure*}
\centering
\includegraphics[width=\textwidth]{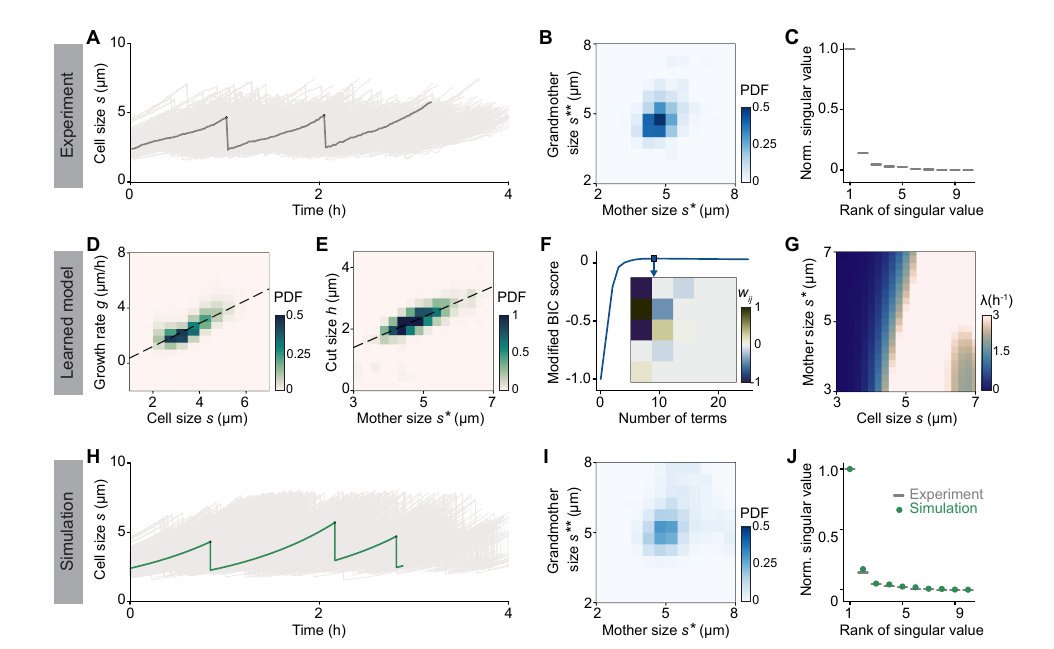}
\caption{The growth and division model learned from experimental data in \textit{B.~subtilis} \cite{nordholt2020biphasic} shows a weak memory of mother size in cell division. 
(\textit{A}) 2643 experimental trajectories of cell sizes are shown in light gray. A typical trajectory is highlighted in dark gray. The doubling time is approximately 1~h. 
(\textit{B}) The joint probability distribution of mother size $\cs{}{\ast}$ (cell size at last division) and grandmother size $s^{\ast\ast}$ (cell size at the second last division) shows that two consecutive division sizes are almost independent.
(\textit{C}) The 10 largest singular values of the joint probability distribution in \textit{B}. The singular values, except for the first one, are relatively small, indicating that cells lack a significant memory of their mother sizes during divisions. 
(\textit{D}) Heatmap of cell size $s$ versus growth rate $g$ from input data. The dashed line shows the learned growth rate $g(s)$. 
(\textit{E}) Heatmap of mother size $s^{\ast}$ versus cut size $h$ from input data. The dashed line shows the learned cut size $h(\cs{}{\ast})$.
(\textit{F}) Modified Bayesian information criterion (BIC) scores of models with varying number of terms. The square marker indicates the selected model with the highest BIC score, whose coefficients $\mathbf{w}$ are shown in the inset.
(\textit{G}) The learned division rate $\lambda$ corresponding to the selected model in \textit{F}.
(\textit{H}) -- (\textit{J}) Simulation results of the selected learned model described by \textit{D} -- \textit{G}. Plots correspond to \textit{A} -- \textit{C}, respectively. 
}
\label{SIfig_bacillus}
\end{figure*}

\clearpage

\begin{figure}
\centering
\includegraphics[width=\textwidth]{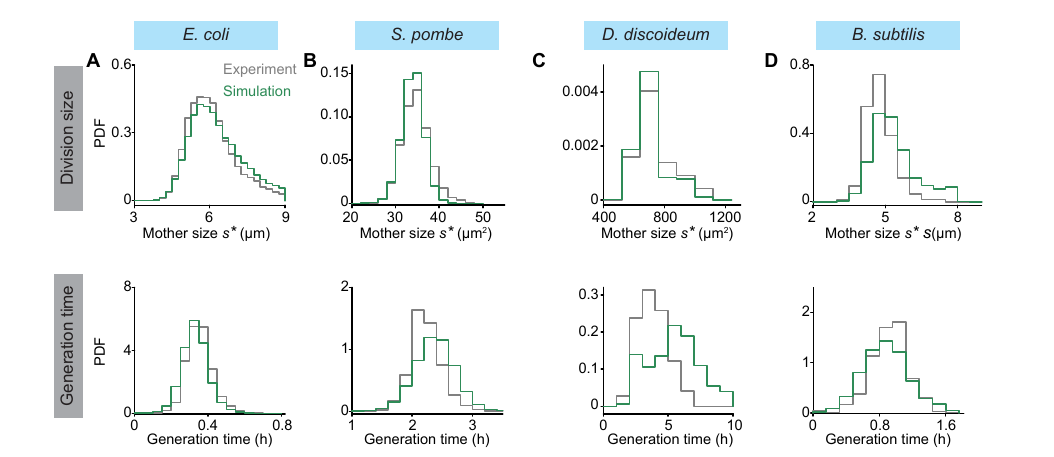}
\caption{Our framework learns cell growth and division models that capture the experimentally measured statistics.
Histograms of cell size at division (\textit{top}) and generation time (\textit{bottom}) are compared between the experiments (gray) and the simulations of the learned models (green) for
(\textit{A}) \textit{E.~coli} \cite{wang2010robust},  (\textit{B})\textit{S.~pombe} \cite{nakaoka2017aging}, (\textit{C}) \textit{D. discoideum} \cite{tanaka2021dynamic}, and (\textit{D}) \textit{B. subtilis} \cite{nordholt2020biphasic}.}
\label{SIfig_statistics}
\end{figure}

\clearpage

\begin{figure}
\centering
\includegraphics[width=\textwidth]{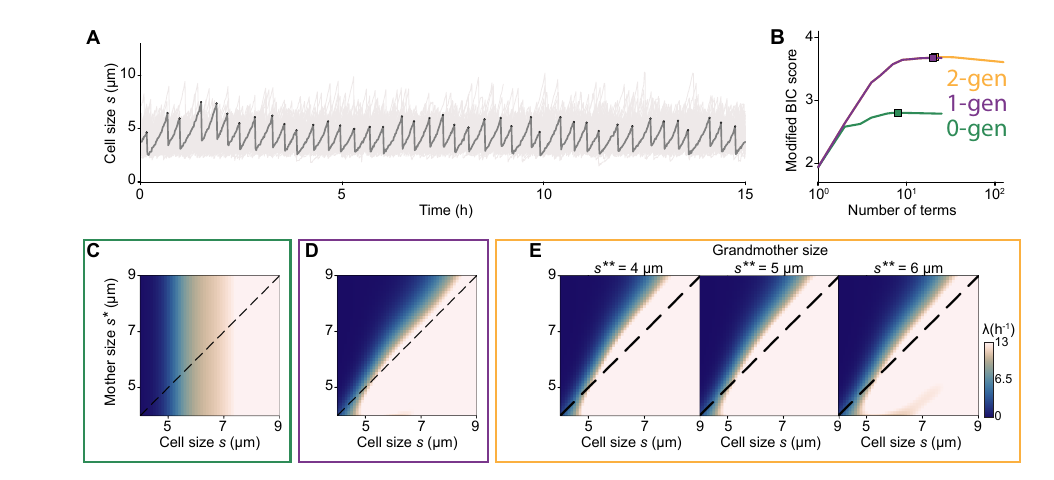}
\caption{
Learning models with different generations of memory from \textit{E.~coli} data \cite{wang2010robust} shows that an one-generation memory model is necessary and sufficient to fit the data.
(\textit{A}) Trajectories of cell sizes from 265 independent \textit{E.~coli} experiments are shown in light gray lines and a
typical trajectory is highlighted in dark gray.
(\textit{B}--\textit{E}) Modified Bayesian information criterion (BIC) scores of models with varying number of terms for a no-memory model (0-gen, green), an one-generation memory model (1-gen, purple), and a two-generation memory model (2-gen, yellow). See text for details. The square markers indicate the selected models with the highest BIC score, for which the learned division rate $\lambda$ are shown in \textit{C} -- \textit{E}. Black dashed line indicates $s = s^*$. (\textit{C}) 0-generation memory model. (\textit{D}) 1-generation memory model. (\textit{E}) 2-generation memory model. 
}
\label{SIfig_012gen_memory}
\end{figure}

\clearpage

\begin{figure*}
\centering
\includegraphics[width=.9\textwidth]{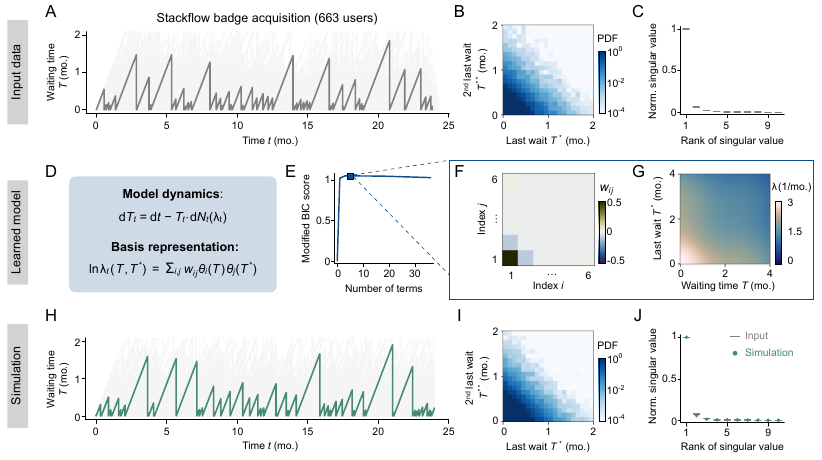}
\caption{The Bayesian inference framework is applicable to learning stochastic dynamics from the Stack Overflow dataset.
This dataset contains the awards history of 663 users in an online question-answering website. Each sequence represents the time points (over a two-year span) when a user receives a badge. We model the waiting time $T$ from the previous badge acquisition using an SDE model with inhomogeneous Poisson noise (see \textit{D}), similar to the one describing cell growth and division dynamics. Our inference framework can then be directly applied to learning the Poisson intensity.
(\textit{A}) Trajectories of waiting time $T$ are shown in light gray and a typical trajectory is highlighted in dark gray.
(\textit{B}) Joint probability distribution of the last wait $T^{\ast}$ (analogous to mother size in the cell division model) and the second last wait $T^{\ast}$ (analogous to grandmother size in the cell division model). 
(\textit{C}) The top 10 singular values (normalized by the largest one) of the joint probability distribution in \textit{B}.
(\textit{D}) We describe the stochastic dynamics using the designated SDE, and we use basis-function methods to approximate $\ln \lambda$.
(\textit{E}) Modified Bayesian information criterion (BIC) scores of models with varying number of terms. The square marker indicates the selected model with the highest BIC score.
(\textit{F},\textit{G}) The learned model coefficients $\mathbf{w}$ (\textit{F}) and division rate $\lambda$ (\textit{G}) corresponding to the selected model in \textit{E}.
(\textit{H}) -- (\textit{J}) Simulation results of the selected learned model described by \textit{D} -- \textit{G}. Plots correspond to \textit{A} -- \textit{C}, respectively. 
}
\label{SIfig_stackexchange}
\end{figure*}

\clearpage

\begin{figure*}
\centering
\includegraphics[width=.9\textwidth]{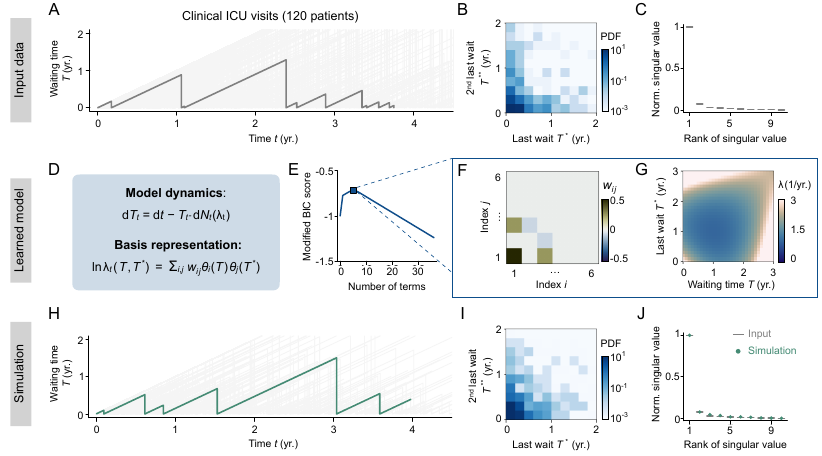}
\caption{The Bayesian inference framework is applicable to learning stochastic dynamics from a medical records dataset. This dataset contains the clinical visit history of 120 patients in an Intensive Care Unit (ICU). Plots correspond to Fig.~\ref{SIfig_stackexchange}.
}
\label{SIfig_icu}
\end{figure*}

\clearpage

\begin{figure*}
\centering
\includegraphics[width=.9\textwidth]{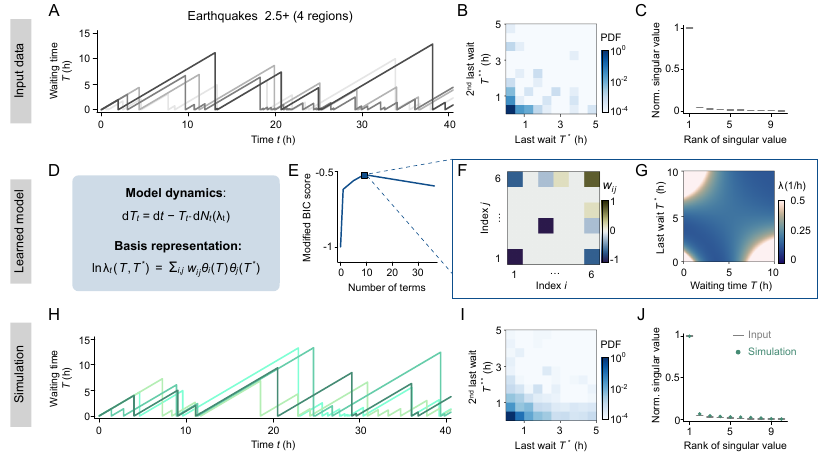}
\caption{The Bayesian inference framework is applicable to learning stochastic dynamics from an earthquake dataset. The dataset records earthquakes of magnitude 2.5 or greater in four U.S. regions near (1) Alaska (longitude [-160$^\circ$, -140$^\circ$] latitude [50$^\circ$, 72$^\circ$]), (2) Hawaii (longitude [-170$^\circ$, -140$^\circ$] latitude [10$^\circ$, 30$^\circ$]), (3) Puerto Rico (longitude [-80$^\circ$, -50$^\circ$] latitude [10$^\circ$, 25$^\circ$]), and (4) California (longitude [-125$^\circ$, -110$^\circ$] latitude [27$^\circ$, 45$^\circ$]), corresponding to dark to light colors in panels \textit{A} and \textit{H}. Plots correspond to Fig.~\ref{SIfig_stackexchange} and \ref{SIfig_icu}.
}
\label{SIfig_earthquake}
\end{figure*}

\clearpage

\begin{figure*}
\centering
\includegraphics[width=.7\textwidth]{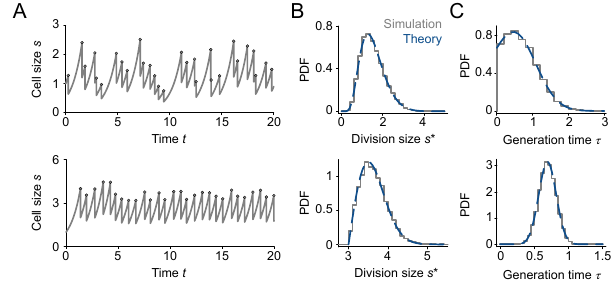}
\caption{Analysis of models with no memory in cell division. 
(\textit{A}) Representative trajectories of cell size from simulations of no-memory models.
(\textit{B} -- \textit{C}) Probability density function (PDF) of (\textit{B})) cell size at division and (\textit{C}) generation time for the corresponding models in \textit{A}. Gray histograms denote simulation results, and blue dashed curve denote theoretical results in Sec.~2\ref{sec:memoryless_analysis}.
The \textit{top} row shows results for $\lambda(\cs{t}{})=\cs{t}{2}$. The \textit{bottom} row shows results for $\lambda(\cs{t}{}) = \alpha \cs{t}{}(\cs{t}{}-\cs{\mathrm{c}}{})$ when $\cs{t}{}>\cs{\mathrm{c}}{}$ and $\lambda(\cs{t}{})=0$ otherwise, where $\alpha=4$ and $\cs{\mathrm{c}}{}=3$.
}
\label{SIfig_nomemory}
\end{figure*}

\clearpage

\begin{figure*}
\centering
\includegraphics[width=.7\textwidth]{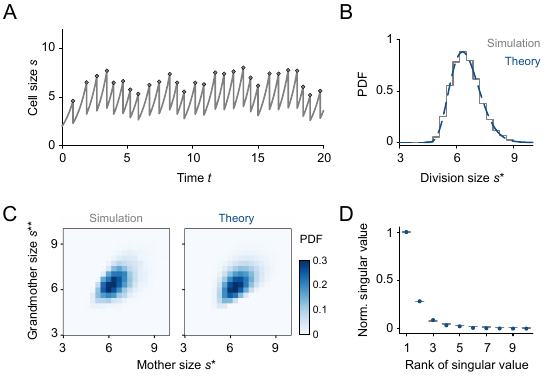}
\caption{Analysis of a model with one-generation memory in cell division. 
(\textit{A}) A representative trajectory of cell size from simulations of a one-generation memory model, where the cell division rate $\lambda(\cs{t}{},\cs{t}{\ast})$ is given by Eq.~\ref{eqn:lambda_analysis_memory}. The simulation parameters are: $\phi=0.5, \cs{\mathrm{c}}{}=4$, and $\alpha = 1$.
(\textit{B}) The probability density function (PDF) of the cell size at division $\cs{}{\ast}$ for the simulation in \textit{A}. Gray histograms denote simulation results, and blue dashed curve denote theoretical results.
(\textit{C}) The joint probability distributions of cell sizes $\cs{}{\ast}$ and $s^{\ast\ast}$ at two consecutive divisions generated by simulation (\textit{left}) and theory (\textit{right}).
(\textit{D}) 
The 10 largest singular values of the joint probability distributions in \textit{C} show good agreement between simulation (gray) and theory (blue).
}
\label{SIfig_analysis_memory}
\end{figure*}

\clearpage

\begin{figure*}
\centering
\includegraphics[width=.7\textwidth]{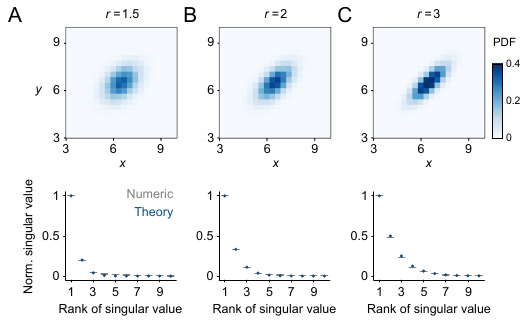}
\caption{Analysis of the singular value spectra for 2D normal distributions. 
The joint probability distributions (\textit{top}), given by Eq.~\ref{eqn:multivariate_gaussian}, and their singular value spectra (\textit{bottom}) are shown for three different parameters: (\textit{A}) $r=1.5$, (\textit{B}) $r=2$, and (\textit{C}) $r=3$. See Sec.~2\ref{sec:svd_analysis} for details. We set $\sigma_+^2 + \sigma_-^2 = 2$ and varied $r=\sigma_+/\sigma_-$.
}
\label{SIfig_analysis_svd}
\end{figure*}

\end{document}